\documentclass[manuscript]{acmart}

\AtBeginDocument{%
  }

\setcopyright{acmlicensed}
\copyrightyear{2018}
\acmYear{2018}
\acmDOI{XXXXXXX.XXXXXXX}

\acmConference[Conference acronym 'XX]{Make sure to enter the correct
  conference title from your rights confirmation emai}{June 03--05,
  2018}{Woodstock, NY}
\acmISBN{978-1-4503-XXXX-X/18/06}

\usepackage{tabularray}
\usepackage{ulem}
\usepackage{tcolorbox}

\begin{document}

\title{An Empirical Study on the Code Refactoring Capability of Large Language Models}

\author{Jonathan Cordeiro}
\email{19jac16@queensu.ca}
\orcid{1234-5678-9012}
\affiliation{%
  \institution{Queen's University}
  \city{Kingston}
  \state{Ontario}
  \country{Canada}
}

\author{Shayan Noei}
\email{s.noei@queensu.ca}
\orcid{0000-0002-5675-7817}
\affiliation{%
  \institution{Queen's University}
  \city{Kingston}
  \country{Canada}}
\email{}

\author{Ying Zou}
\email{ying.zou@queensu.ca}
\orcid{0000-0002-5335-0261}
\affiliation{%
  \institution{Queen's University}
  \city{Kingston}
  \country{Canada}
}

\renewcommand{\shortauthors}{Trovato et al.}

\begin{abstract}
Large Language Models (LLMs) aim to generate and understand human-like text by leveraging deep learning and natural language processing techniques. 
In software development, LLMs can enhance the coding experience through coding automation, reducing development time and improving code quality. Code refactoring is a technique used to enhance the internal quality of the code base without altering its external functionalities. Leveraging LLMs for code refactoring can help developers improve code quality with minimal effort.
In this paper, we conduct an empirical study to assess the quality of the refactored code generated by StarCoder2, which is an LLM designed for code generation. Specifically, we (1) evaluate whether the code refactored by the LLM or by developers is more effective at improving code quality, (2) understand the differences between the types of refactoring applied by the LLM and developers and compare their effectiveness, and (3) evaluate whether the quality of the refactored code generated by the LLM can be improved through one-shot prompting and chain-of-thought prompting.
We analyze the refactoring capabilities of StarCoder2 and developers on 30 open-source Java projects. We find that StarCoder2 reduces code smells by 20.1\% more than developers on automatically generated refactorings. StarCoder2 excels in reducing more types of code smells, such as Long Statement, Magic Number, Empty Catch Clause, and Long Identifier. Developers perform better in fixing complex issues, such as Broken Modularization, Deficient Encapsulation, and Multifaceted Abstraction. Furthermore, StarCoder2 outperforms developers in refactoring types that are more systematic and repetitive. However, developers surpass in refactorings that require a deeper understanding of code context and architecture. Our findings show that One-shot prompting improves unit test pass rate over the zero-shot prompt by 6.15\% and reduces code smells at a 3.52\% higher rate. When generating five refactorings per input, StarCoder2 achieves a unit test pass rate of 28.8\% higher than when generating one refactoring per input, which indicates that the combination of one-shot prompting with multiple refactoring generations per input leads to the best performance. By providing insights into the capabilities and best practices for integrating LLMs like StarCoder2 into the software development process, our study aims to enhance the effectiveness and efficiency of code refactoring in real-world applications.
\end{abstract} 




\maketitle

\section{Introduction}
\label{sec:Introduction}
Automatic code generation allows developers to produce code more efficiently and consistently~\cite{svyatkovskiy2020intellicode}. The advancement of Large Language Models (LLMs) has demonstrated its ability in generating code snippets and solving complex programming problems~\cite{fan2023large, wei2023copiloting}. LLMs are trained using an enormous amount of open-source code; therefore, if the training code has flaws, the generated code could suffer from poor design~\cite{10.1145/3487569} or introduce technical debts~\cite{chang2024survey}. Therefore, the integration of the generated code into the software development process may increase maintenance costs and reduce the overall quality of the software~\cite{10.1145/3487569}.  

Code refactoring is the process of enhancing the internal quality of the code without altering its external behavior~\cite{fowler1999refactoring, fowler2018refactoring, noei2023empirical}. Refactoring types are the specific techniques used, such as renaming variables for clarity, extracting methods to reduce complexity, and moving classes or methods~\cite{Refactoring.Guru}. Refactoring is typically associated with the elimination of code smells~\cite{10.1145/3180155.3182532} which are design flaws that violate design principles and compromise the maintainability~\cite{van2002java}. As a result, the refactored code can improve software design, make code easier to understand, and remove dependencies among software components~\cite{1265817, article, noei2023empirical}. Human developers bring contextual knowledge and experience to refactoring, which is challenging for automated systems to replicate~\cite{van2002java, fowler1999refactoring, fowler2018refactoring}. Recent studies have explored the potential of LLMs (e.g., GPT-3.~\cite{10479398}) in automating code refactoring, where models are provided with a prompt and the code to be refactored~\cite{10479398, Choi2024aa}. However, there are no systematic studies to understand the impact of LLM-generated refactorings on code quality improvement~\cite{10.1145/3643661.3643953}. 

To have a better understanding of how LLMs manage refactoring tasks without compromising the functionality of the code, we conduct an empirical study to evaluate the effectiveness of LLMs in refactoring tasks and compare their performance against that of human developers. We aim to provide an evaluation framework to determine how well LLMs can replicate or surpass human expertise in code refactoring tasks. Given the wide range of refactoring types that developers can perform, we strive to gain insight into the coverage of refactoring types that LLMs can perform, compared to those performed by developers. Moreover, we explore the possibility of improving the capabilities of LLMs in refactoring by applying one-shot and chain-of-thought prompting to enhance LLMs' performance in refactoring.

The LLMs are trained on public data, including open-source repositories~\cite{xu2022systematic}. Therefore, if the LLMs are trained using refactored code from the developers that we aim to compare LLM-generated refactored code with, it cannot truly reflect LLMs' refactoring capabilities. Overfitting could occur if the LLM has been trained on the same code, potentially leading to artificial inflation of its performance and resulting in over-optimistic results~\cite{Brownlee_2020}. Popular LLMs such as OpenAI's GPT-4~\cite{openai2024gpt4}, Google's PaLM~\cite{chowdhery2022palmscalinglanguagemodeling}, and Meta's LLaMA~\cite{touvron2023llamaopenefficientfoundation} do not provide their training datasets publicly. To mitigate the risk of data leakage, we choose StarCoder2-15B-instruct~\cite{lozhkov2024starcoder}, trained on the publicly available dataset, The Stack v2~\cite{lozhkov2024starcoder}. Therefore, we ensure that our selection of open-source projects for evaluating the LLM has not been included in the LLM's training dataset.

StarCoder2 has exceptional performance on the HumanEval benchmark~\cite{10.1145/3580305.3599790} which assesses the model's ability to generate functional code for a variety of programming tasks, making it a strong indicator of the model's effectiveness in real-world coding scenarios. StarCoder2 achieves 46.3\% for the pass@1 metric on the HumanEval benchmark, meaning that 46.3\% of the generated code snippets perform the required task correctly on the first attempt. 

In this paper, we conduct an empirical study using 30 open-source Java projects, that are not included in the Stack-v2~\cite{li2023starcoder}. We extract 5,194 refactoring commits from the 30 projects. To compare the refactorings conducted by human developers and the ones generated by StarCoder2, we leverage the rich commit history available in the subject projects. We aim to extract the code in a file level before a developer’s refactoring and the code after the developer’s refactoring. Then we use prompt engineering to instruct StarCoder2 to refactor the code before a developer conducts the refactoring on the same code. We compare the refactorings produced by StarCoder2 and the ones by developers by analyzing the reduction of code smells, as well as the impact of different refactoring types on code quality improvement. We aim to investigate the following four research questions:

\textbf{RQ1: Can LLMs outperform developers in code refactoring?}
We aim to assess whether StarCoder2 can be used as a reliable solution to automate code refactoring. We compare the distribution of the refactoring operations performed by StarCoder2 and developers. We evaluate their effectiveness with the reduction of code smells and their improvement on code quality measured by various code metrics.  We observe that StarCoder2 achieves a significantly higher performance in reducing code smells by 44.36\%, compared to a 24.27\% reduction rate for the developer. StarCoder2 excels in improving code quality measured by code metrics, often surpassing developers in these areas.

\textbf{RQ2: Which types of code smells are most effectively reduced by LLMs or developers?}
To measure the quality improvement capabilities of StarCoder2 in refactoring, we identify the types of code smells that can be effectively removed by StarCoder2 or developers. We observe that StarCoder2 outperforms developers to address systematic and repetitive issues, such as condensing a long statement or shortening a long parameter list. However, developers show superiority in handling complex, context-dependent code smells, such as correcting a broken modularization or deficient encapsulation.

\textbf{RQ3: Which refactoring types are most effective for improving code quality?}
To understand the capabilities of StarCoder2 in handling different types of refactorings, we compare the refactoring types that it performs to effectively eliminate code smells and improve code quality with those performed by developers. Our findings show that StarCoder2 is particularly effective in refactoring types that improve code within a class. Developers on the other hand excel in refactoring types that involve changes affecting multiple classes.

\textbf{RQ4: How does prompt engineering affect the quality of LLM-generated refactorings?}
To explore the impacts of prompt engineering to improve the capabilities of StarCoder2, we use chain-of-thought and one-shot prompting techniques. The results show that the one-shot prompting yields the highest unit test pass rate of 34.51\%, marking an improvement of 6.15\% over zero-shot prompting, and a smell reduction rate (SRR) of 42.97\%, which is an increase over the zero-shot prompt by 3.52\%. The chain-of-thought prompt, where we give the LLM refactoring type suggestions along with a definition, achieves a 32.22\% unit test pass rate and a 42.34\% SRR, which improves upon the numbers from the zero-shot prompting by 3.86\% and 2.89\% respectively. 

Our work makes the following main contributions:
\begin{itemize}
\item We conduct a comprehensive evaluation of the capabilities of StarCoder2 in automated code refactoring.
\item We develop an evaluation framework for measuring the effectiveness of LLMs in code refactoring, with a focus on code quality improvement and functionality preservation.
\item We compare different prompting techniques, including chain-of-thought and one-shot prompting, to assess their impact on the generation of high-quality refactorings and offer practical guidance on their application.
\item We provide a replication package to enable the reproducibility of our study. The replication package of the study can be accessed at: \url{https://github.com/Software-Evolution-Analytics-Lab-SEAL/LLM_Refactoring_Evaluation}
\end{itemize}

\textbf{Paper Organization.} The remainder of our study is organized as follows. Section~\ref{sec:experiment_setup} describes the experiment setup of this study. Section~\ref{sec:results} presents the motivation, approaches, and results of our research questions. Section~\ref{sec:threads_to_validity} discusses the threats to the validity of our findings. Section~\ref{sec:related_work} surveys related studies and compares them to our work. Finally, we conclude our paper and present future research directions in Section~\ref{sec:conclusion}.
\section{Experiment Setup}
\label{sec:experiment_setup}
\begin{figure*}[t]
\centering
\includegraphics[width=\linewidth]{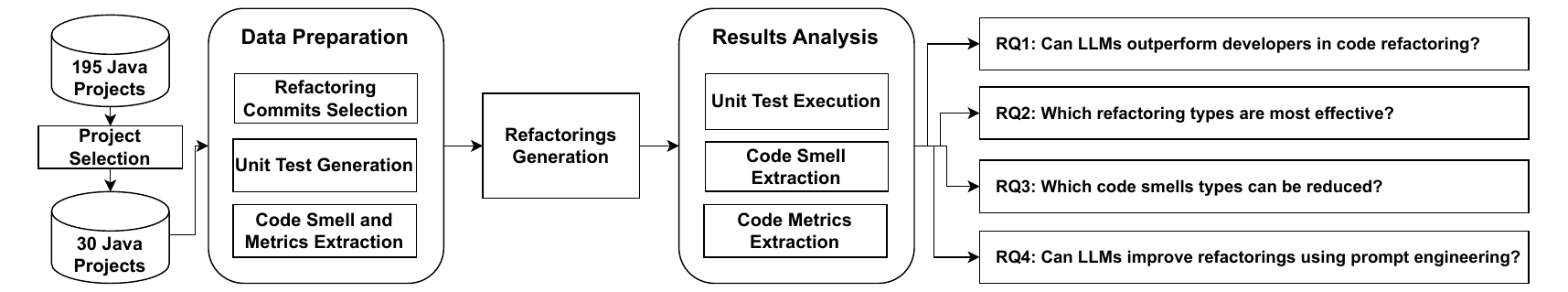}
\caption{Overview of Our Approach for Data Collection and Answering Research Questions.}
\label{fig:approach_overview}
\end{figure*}

We start with selecting projects that are not included in the training dataset of StarCoder2. Then we conduct the data processing to extract the refactoring-related commits, associated files containing refactorings, code smells detection, and code metrics measurement. Next, we prompt StartCoder2 to generate refactorings on each extracted file in a commit before the developers' refactoring operations. Furthermore, we analyze the refactored code generated by StarCoder2 and answer the three research questions. The rest of the section presents the experiment of our study, including our data collection and data analysis approaches.

\subsection{Project Selection}
\label{sec:project_selection}
We conduct our study using the 20-MAD dataset~\cite{claes2020mad}, which includes 765 Apache projects along with their historical development information, such as code commit information. We limit our study to projects primarily written in Java. Selecting projects using Java is based on the strong performance of refactoring detection tools developed for Java~\cite{Tsantalis:ICSE:2018:RefactoringMiner, tsantalis2020refactoringminer} and Java’s popularity~\cite{berkeley2020indemand, TIOBE_2022} among developers~\cite{noei2023empirical}, which aligns with our goal of examining refactorings in a language where refactoring practices are well-established and widely adopted.

To include projects with sufficient historical development and tool support, inspired by previous work~\cite{noei2023empirical}, we exclude the projects that: (1) have less than 80\% of Java source code; (2) have less than the 1\textsuperscript{st} quantile of commit counts (\textit{i.e.}, $<$ 1,021 commits); (3) have a short lifespan (\textit{i.e.}, $<$ one-year of commit history); and (4) projects with less than one-year of commit history as projects with a longer lifespan and more commits are likely to have more refactoring operations~\cite{fowler1999refactoring}. As a result, we obtain 195 Java projects. 

To avoid data leakage, we filter out the 135 Java projects from our initial dataset that are included in the StarCoder2. To ensure a balanced distribution of refactoring commits across projects, we filter out projects with fewer than the median number (\textit{i.e.}, 129) of refactoring commits in a final 30 Java projects for our analysis. To obtain the refactoring information for each commit in the selected projects, we utilize the dataset provided by a previous study \cite{noei2023empirical}, which is comprised of commits from the 195 Java Apache projects mentioned above. We use the commits of the 60 projects from this dataset for our analysis. The commits contain refactorings detected by RMiner~\cite{Tsantalis:ICSE:2018:RefactoringMiner}, and include 59 different types of refactorings along with the modified lines of code, which is used to calculate code churn. Code churn refers to the sum of lines added and removed during a commit~\cite{fowler1999refactoring}. The evaluation of RMiner on this dataset performed in previous work~\cite{noei2023empirical}, shows an overall precision of 99.7\% and a recall of 94.2\%, confirming it as an effective refactoring detection tool~\cite{Tsantalis:ICSE:2018:RefactoringMiner}.

\subsection{Data Processing}
\label{sec:Dataset}
\subsubsection{Refactoring Commits Selection}
\label{sec:commits_selection}
To ensure that the developer performs the same task (\textit{i.e.,} refactoring) as the StarCoder2, rather than other tasks such as development or bug fixing, we select commits that 100\% of the lines of code churn consist of refactoring operations. Refactoring code churn refers to the sum of lines added and removed in the process of a refactoring operation. A refactoring commit is identified by comparing the refactoring code churn with the total code churn within a commit shown in the following equation~\cite{noei2023empirical}:

\begin{equation}
   \text{Refactoring Ratio of Commit (\textit{i})} = \frac{\text{Refactoring Code Churn of Commit (\textit{i})}}{\text{Total Code Churn of Commit (\textit{i})}} \times 100\%
   \label{equ:code_churn_ratio}
\end{equation}

If the code churn is entirely due to refactoring (\textit{i.e.}, Ratio = 100\%), then the commit is classified as a refactoring commit.

\subsubsection{Code Smell Extraction}
Code smells are indicators of underlying design or implementation issues that may affect the quality, such as the maintainability and readability of the software~\cite{noei2023empirical}. Code refactoring is commonly associated with the elimination of code smells~\cite{10.1145/2950290.2950305, 10.1145/3106237.3106259, tsantalis2020refactoringminer}. We use DesigniteJava 2.5.2~\cite{sharma2016designite} to extract code smells by inputting the source code of each Java file from before and after each refactoring commit. DesigniteJava offers comprehensive coverage in detecting code smells across multiple categories~\cite{10.1145/3643991.3644881}. It can detect 46 different types of code smells, including 7 architecture smells, 18 design smells, 9 implementation smells, 4 testability smells, and 8 test smells.
\begin{itemize}
    \item Architecture smells describe the characteristics in the system's architecture that indicate potential issues affecting the overall software quality. For example, unstable dependencies code smell describes the case that relies on frequently changing modules.
    \item Design smells indicate poor adherence to design principles that can negatively impact the software's modularity, flexibility, and reusability. These smells suggest that the design may not support future changes or scale effectively, leading to increased technical debts and higher maintenance costs. For example, unnecessary abstraction describes the code smell with too many layers or indirections that could obscure the code's intent.
    \item Implementation smells are signs in the source code to flag problematic code that could make the code harder to maintain, understand, or extend. Common examples include large classes, repeated code, or very long methods, which indicate that the code may need improvement to enhance its quality and ease of maintenance.
    \item Testability smells refer to the degree to which the development of test cases can be facilitated by the software design choices. Examples include hard-wired dependencies and excessive dependency.
    \item Test smells are resilted from bad programming practices in unit test code indicating potential design problems in the test code. Examples include empty test, unknown test, and constructor initialization.
\end{itemize}

\subsubsection{Code Metrics Computation}
Code metrics provide valuable insights into the quality of the code, such as maintainability, and complexity~\cite{nunez2017source}. We collect code metrics that are used to evaluate complexity, cohesion, coupling, and modularity. The description of the collected code metrics is listed in Table \ref{tab:metrics}. Modularity metrics help assess the degree to which the system is divided into independent components, which promotes ease of maintenance and testing. Coupling metrics indicate the degree of interdependence between classes, where lower coupling is preferred for better modularity and flexibility. Cohesion metrics measure how closely related the responsibilities of a class are, with higher cohesion typically leading to improved code clarity and design quality~\cite{10.1145/3408302}. We use the Understand tool~\cite{understand_software}, which is a static code analysis tool to extract code metrics.

\begin{table}
\centering
\scriptsize
\caption{Description of Metrics Used in the Study~\cite{understandmetrics}.}
\label{tab:metrics}
\resizebox{\linewidth}{!}{%
\begin{tblr}{
  width = \linewidth,
  colspec = {Q[150]Q[304]Q[92]Q[446]},
  hline{1,13} = {-}{0.08em},
  hline{2} = {-}{0.05em},
}
\textbf{Metric}                      & \textbf{Description}                                                       & \textbf{Quality Attribute} & \textbf{Rationale for Inclusion}                                                                                    \\
{Count Class \\Coupled}              & Number of other classes coupled to.                   & Coupling                   & Helps identify tightly coupled classes, indicating potential areas for refactoring to reduce complexity.            \\
{Count Class\\Coupled Modified}      & Number of other non-standard classes coupled to.       & Coupling                   & Focuses on the complexity introduced or modified during changes, relevant for assessing refactoring impact.         \\
{Count Class\\Derived}               & Number of immediate subclasses.                       & Modularity                 & High values may indicate a deep inheritance hierarchy, affecting maintainability and ease of modification.          \\
{Count Decl\\Class Variable}         & The number of class-level variables declared in a class.                   & Modularity                 & A large number of class variables can reduce cohesion and increase the potential for errors.                        \\
{Count Decl\\Instance Variable}      & The number of instance variables declared in a class.                      & Modularity                 & High instance variable counts may signal a class doing too much, reducing its maintainability.                      \\
{Percent Lack\\of Cohesion}          & The percentage of methods in a class that do not share instance variables. & Cohesion                   & Indicates the degree of cohesion in a class, with higher values suggesting poor design and low cohesion.            \\
{Percent Lack of\\Cohesion Modified} & A variation of PercentLackOfCohesion, modified for accessor methods.        & Cohesion                   & Assesses the impact of changes on class cohesion, useful for understanding the effects of refactoring.              \\
Avg Cyclomatic                       & The average cyclomatic complexity for all nested functions or methods.    & Complexity                 & Provides an overall view of the code's complexity, helping to identify areas that may need simplification.          \\
Cyclomatic                           & The McCabe cyclomatic complexity of a single method or function.                  & Complexity                 & Directly measures the complexity of a method, indicating potential difficulties in testing and understanding.       \\
Max Cyclomatic                       & The maximum cyclomatic complexity of any method in a class or project.     & Complexity                 & Highlights the most complex methods, which could be refactoring targets to improve maintainability.                 \\
Sum Cyclomatic                       & The total cyclomatic complexity of all methods in a class or project.      & Complexity                 & Summarizes the overall complexity, aiding in assessing the overall maintainability and testability of the codebase. 
\end{tblr}
}
\end{table}

\subsubsection{Tests Preparation}
\label{sec:tests_preparation}
Code refactoring should not alter the external functionality of the code~\cite{fowler1999refactoring}. To evaluate whether the refactorings generated by StarCoder2 retain their original functionality, we execute unit tests on the refactored code. As some projects don't have test cases, we cannot obtain the original test cases from all refactoring commits (identified in Section \ref{sec:commits_selection}). Therefore, we generate test cases for all the refactoring commits to test whether the refactored code alters functionalities or introduces new defects. We utilize EvoSuite~\cite{inproceedings}, which automatically generates test suites that are optimized to maximize code coverage across multiple criteria, including line coverage (\textit{i.e.}, ensuring all lines of code are executed), branch coverage (\textit{i.e.}, testing each possible path in the code), and output coverage (\textit{i.e.}, verifying the correctness of program outputs). EvoSuite incorporates JUnit 4 assertions, which are used to capture and validate the current behavior of the tested classes. We take the following steps to generate test cases using EvoSuite:
\begin{enumerate}
    \item \textbf{Create a project for the code extracted from each refactoring commit:} For each refactoring commit, we set up a project that contains only the files modified by the commit in their state before the refactoring was performed. We ensure the project is configured with the necessary dependencies, either by extracting the pom.xml from the repository or generating one based on a template that includes essential libraries (\textit{e.g.}, JUnit and Mockito) for unit testing. A pom.xml file is an XML configuration file used by Apache Maven to manage project dependencies, build configurations, and other essential project information. This ensures that the project can be compiled and executed for further analysis. Since only files changed in the commit are included, some dependencies may be missing. To handle this, Mockito~\cite{mockitoMockitoFramework} is used to create mock objects for unavailable dependencies, allowing tests to execute effectively. A bash script, included in our replication package, retrieves or creates the pom.xml and builds the Java project using Apache Maven~\cite{miller2010apache}. EvoSuite then builds the test cases based on the pom.xml.
    \item \textbf{Compile the project:} After setting up the project for each commit, we compile the code using Apache Maven to ensure that it builds successfully. Compiling the project verifies that all dependencies are correctly set up and that the source code is free of compilation errors, which is crucial before generating test cases.
    \item \textbf{Generate test cases with EvoSuite:} Once the project is compiled, we use EvoSuite to automatically generate unit tests. EvoSuite is run on each Java class within the project. For every class, EvoSuite generates a suite of tests designed to cover different behaviors and edge cases of the code. These generated tests are stored in a separate directory within each commit folder, providing a structured and isolated location for the test cases.
\end{enumerate}

We manually validate the generated unit tests to ensure their functionality as EvoSuite may run into errors when generating unit tests and some unit tests may be incomplete. We randomly select 381 unit tests from 39,309 generated unit tests to assess each one to determine if it is valid and functional. Of the 381 unit tests, 348 tests (91.3\%) were valid and functional, meaning they executed as expected without encountering the above issues. After this validation, we filter out unit tests that match patterns of failing tests as follows:
\begin{itemize}
    \item Tests that fail due to missing the required dependencies - improper configuration in the initial states.
    \item Overly strict assertions, where minor variations in outputs or system state that do not affect functionality cause the test to fail.
\end{itemize}

\subsection{Hardware Environment}
Our experiment is conducted using 4 Nvidia A100 GPUs on a server, each with 80 GB of memory. The operating system of the server is Ubuntu 22.04.4 LTS. It takes approximately 6 days to generate refactorings for 5,194 commits, and takes around 1.6 minutes per commit. The GPUs are utilized for StarCoder2 model inferencing.
\section{Results}
\label{sec:results}
In this section, we provide the motivation, approach, and findings for each of our research questions.

\subsection{\textbf{RQ1. Can LLMs outperform developers in code refactoring?}}
\subsubsection{\textbf{Motivation}}
Developers typically perform refactoring by manually analyzing and editing code to improve maintainability and readability~\cite{fowler1999refactoring, fowler2018refactoring}. This requires developers to rely on their expertise and knowledge of best practices, which can make the process both time-consuming and prone to errors~\cite{1265817}.
In contrast, LLMs generate refactorings by processing the prompt and code through pre-trained models, using large-scale datasets and patterns learned from vast amounts of code to suggest improvements automatically. 
This research question aims to evaluate whether LLMs, particularly StarCoder2, can outperform developers in refactoring tasks by comparing the quality of refactorings generated by StarCoder2 and developers.

\subsubsection{\textbf{Approach}}
\label{sec:approach_rq1}
To compare refactorings generated by StarCoder2 with those performed by developers, we first select a set of refactoring commits from developers. Next, we generate refactorings using StarCoder2. Finally, we use code quality metrics to compare the quality of the code refactored by developers with that refactored by StarCoder2. Each step of this process is detailed as follows:

\textbf{Commit selection: }To set a foundation for comparison, we establish the developer's refactored code as the baseline. We only use commits with 100\% of the code churn consisting of refactoring operations, ensuring that the changes made by developers are solely for refactoring purposes and do not include any other development activities, such as adding new features. The selected commits represent valid refactoring operations, as they are accepted pull requests and pass the unit tests generated for code before the refactoring. Generating refactorings on all available commits would be computationally expensive and time-consuming. From our 30 projects mentioned in Section \ref{sec:project_selection}, we select a sample size of commits with a confidence level of 95\% and a margin of error of 5\%, ensuring a representative sample from each project~\cite{hazra2017using}, to generate refactorings using StarCoder2. By using a sample size, we can maintain confidence in the evaluation results while balancing the trade-off between computational feasibility and accuracy. Therefore, we have 5,194 commits containing a total of 39,309 files for testing the model.

\textbf{Refactoring generation: }We use the code segment containing the code before refactoring as input to the LLMs, so we can use the generated code from the LLM to compare with the refactored code by the developer on the same snippets of code in a commit. The prompt comprises snippets of code across the files in a commit.

\begin{figure}[h]
    \centering
    \begin{minipage}{0.80\linewidth} 
        \centering
        \fbox{
            \parbox{0.78\textwidth}{ 
                \textbf{Instruction} \\
                You are a powerful model specialized in refactoring Java code. Code refactoring is
                the process of improving the internal structure, readability, and maintainability of a software codebase without altering its external behavior or functionality. You must output a refactored version of the code.

                \vspace{0.3cm}
                \textbf{Prompt} \\
                \textbf{\# unrefactored code snippet(java):} \\
                \{code\_segment\_before\_refactoring\}
                
                \vspace{0.3cm}
                \textbf{\# refactored version of the same code snippet:}
            }
        }
    \end{minipage}
    \caption{Zero-shot Prompt Used to Instruct StarCoder2 to Conduct Refactoring}
    \label{fig:model_prompt}
\end{figure}

Figure \ref{fig:model_prompt} shows the template zero-shot prompt, which is a method where the model receives no additional examples or guidance beyond the initial instruction~\cite{NEURIPS2022_8bb0d291}, used to instruct StarCoder2 to take the developers' code snippet before refactoring as input. We employ the "\#" symbol as a delimiter to ensure the LLM focuses on the instructions even with a block of code in the prompt~\cite{10.1145/3544548.3581388}. The LLM is specifically prompted to refactor Java code and we break the code snippets into chunks (\textit{i.e.}, one chunk for each file in a commit). We create a zero-shot prompt for each refactoring commit using a Python script, which is loaded onto an Nvidia A100 GPU. 

\textbf{Test Pass Rate Evaluation: }We use Pass@1, Pass@3, and Pass@5, all using the same zero-shot prompt, to assess the functional correctness of the refactoring operations:
\begin{itemize}
    \item \textbf{pass@1:} We generate a single refactored solution for each code snippet in a commit. This solution is evaluated by executing the corresponding unit tests. If the refactored solution passes the unit test, it is considered a successful refactoring under the Pass@1 criterion. 
    \item  \textbf{pass@3:} We generate three refactored solutions for each code snippet in a commit. Each solution is independently evaluated by running the corresponding unit tests. If at least one of these three refactored versions passes all unit tests, the refactoring is considered successful under pass@3. 
    \item \textbf{pass@5:} We generate up to five refactored solutions for each code snippet in a commit. As with pass@3, we run the unit tests on each of the five refactored versions. If at least one solution successfully passes all unit tests, the refactoring is deemed successful under pass@5. 
\end{itemize}

In pass@3 and pass@5 scenarios, if multiple refactoring solutions pass the unit test, we select the best solution by analyzing which solution reduces the most amount of code smells. If two or more solutions pass the unit tests and reduce the same amount of code smells, we select the best solution based on which one has the highest improvement in code metrics (\textit{i.e.} cohesion, coupling, modularity, and complexity). 

To assess the effectiveness of StarCoder2 in improving overall code quality after passing unit tests, we use two categories of measurements: code smells and code metrics. For code smell reduction and code quality metrics improvement, we calculate the improvement rate using the following formula. Let \(A_{\text{before}}\) represent the value of the attribute (code smells or code metric) before refactoring, and \(A_{\text{after}}\) represent its value after refactoring. The improvement rate (\textit{IR}) is calculated as follows:
\begin{equation} 
    IR = \frac{A_{\text{before}} - A_{\text{after}}}{A_{\text{before}}} \times 100%
    \label{eq:IR}
\end{equation}

We collect code metrics from three versions of the code using the static analysis tool, Understand~\cite{understand_software}: (1) the unrefactored code, to assess its quality before refactoring; (2) the developer-refactored code, to evaluate its quality after developer refactoring; (3) and the LLM-refactored code, to assess its quality after LLM refactoring. This analysis allows us to systematically compare the effectiveness of refactorings carried out by developers and the LLM respectively. To examine the quality of refactorings produced by the LLM and developers, we examine the following null hypothesis:

\begin{itemize}
    \item {\it \textbf{H\textsubscript{0}:} The refactoring performed by the LLM is as effective as those performed by human developers in reducing code smells and improving code metrics.}
\end{itemize}

We test H\textsubscript{0} by comparing unit test pass rates, code smell reduction rates, and code metric improvement rates between the LLM and developers. We perform a Mann-Whitney U-test~\cite{doi:https://doi.org/10.1002/9780470479216.corpsy0524} on the distributions of code smell reduction rates across the 30 projects between developers and the LLM, using a confident level of 5\% (\textit{i.e.}, p-value<0.05). The U-test assesses whether two or more samples originate from the same distribution. It does not assume a normal distribution since it is a non-parametric statistical test. We also perform the U-test on the distributions of code metric improvement percentages across the 30 projects for the LLM and developers.

We quantify the differences in the effect size using Cliff's delta (\(\delta\))~\cite{meissel2024cliffs}, which measures the degree of overlap between the distributions from StarCoder2 and developers. A higher effect size indicates a larger magnitude of the differences between the two approaches for that particular type of code smell. A Cliff's delta greater than 0.15 indicates a small effect size, between 0.33 and 0.47 indicates a medium effect size, and greater than 0.47 indicates a large effect size~\cite{meissel2024cliffs}.
\label{sec:cliff's delta}

\subsubsection{\textbf{Findings}}
\begin{table}[t]
    \centering
    \scriptsize
    \caption{Comparison of Unit Test Pass Rates and SRR Across the 30 Projects}
    \begin{tabular}{l|c|c|c|c|c}
        \hline
        \textbf{Refactoring} &  & \multicolumn{2}{|c|}{\textbf{Unit Test Pass Rate}} & \multicolumn{2}{|c}{\textbf{SRR}} \\
        & & \textbf{Median} & \textbf{Average} & \textbf{Median} & \textbf{Average} \\ \hline
         & Pass@1 & 26.8\% & 28.4\% & 37.5\% & 39.5\% \\ 
        LLM & Pass@3 & 47.0\% & 48.5\% & 39.6\% & 40.8\% \\ 
         & Pass@5 & \textbf{55.4\%} & \textbf{57.2\%} & \textbf{43.2\%} & 44.4\% \\ \hline
        Developer & - & 100\% & 100\% & 23.5\% & 24.3\% \\ 
        \bottomrule
    \end{tabular}
    \label{tab:unit_test_srr_comparison_starCoder2}
\end{table}

\begin{figure}[h]
    \centering
    \begin{minipage}[t]{0.49\linewidth}
        \centering
        \vspace{0pt} 
        \includegraphics[width=\linewidth]{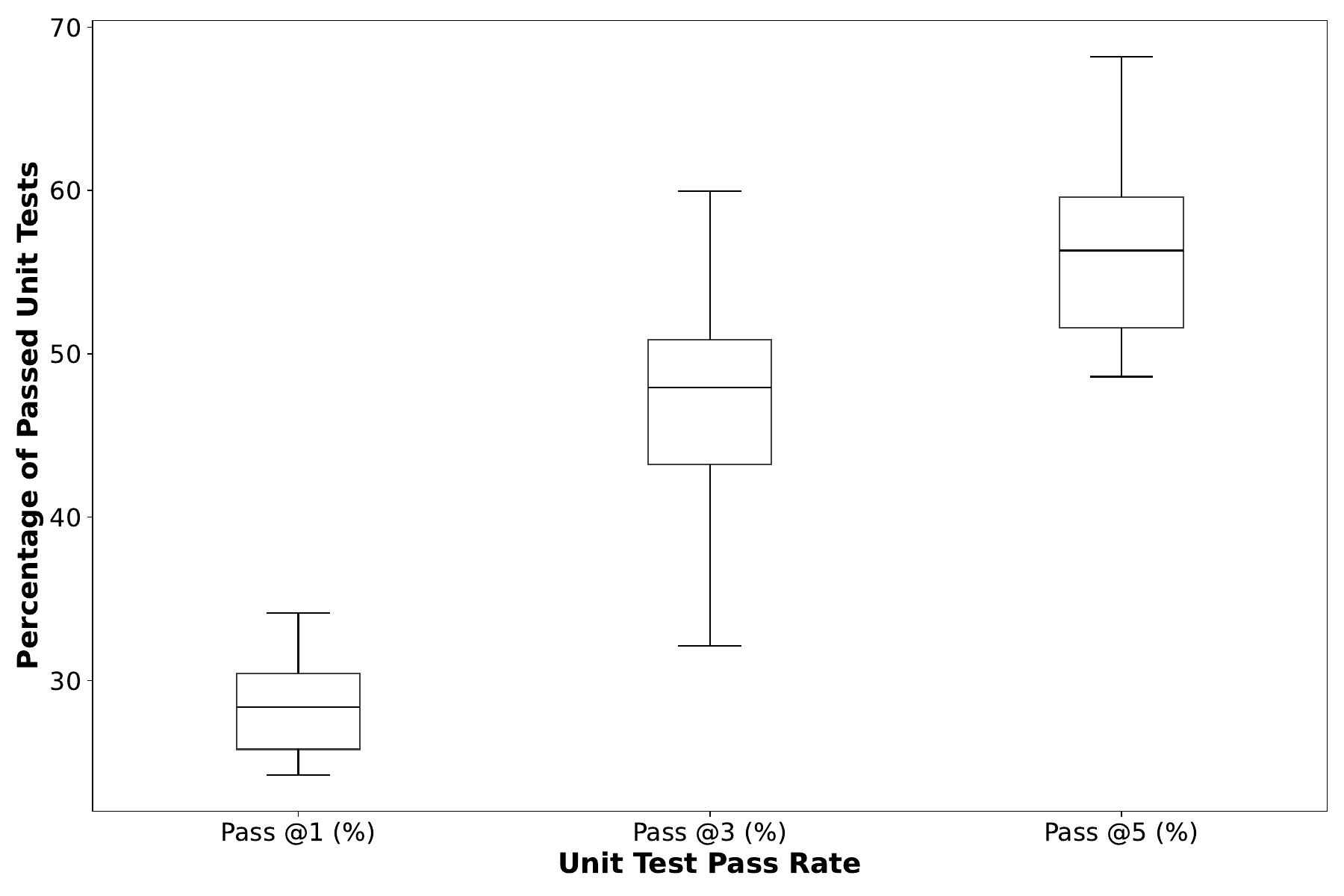}
        \caption{Distribution of Unit Test Pass Rates Across 30 Software Projects After StarCoder2-Generated Refactorings.}
        \label{fig:smells_agg}
    \end{minipage}
    \hfill
    \begin{minipage}[t]{0.49\linewidth}
        \centering
        \vspace{0pt} 
        \includegraphics[width=\linewidth]{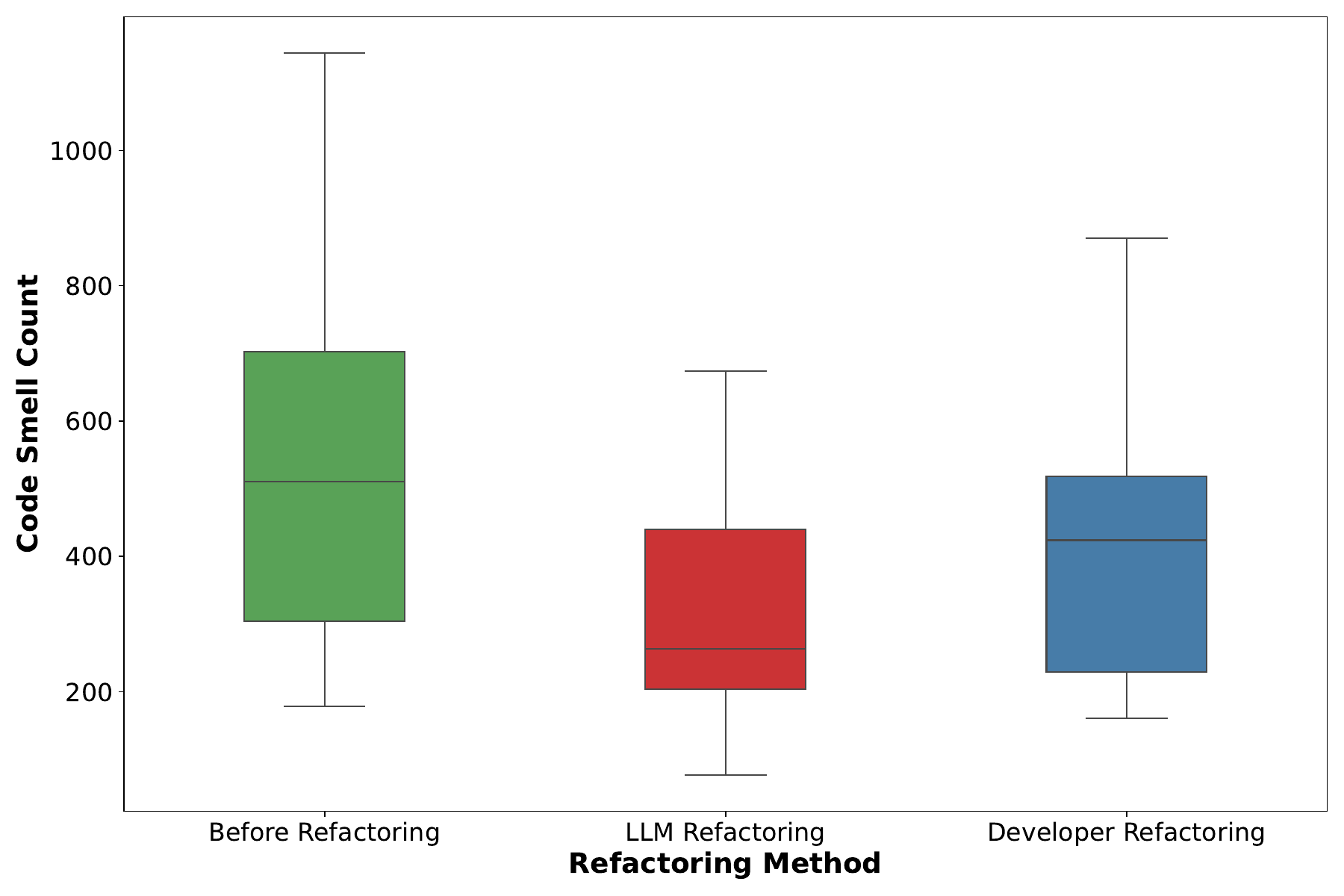}
        \caption{Distribution of Code Smells Across 30 Software Projects After LLM-Generated Refactorings.}
        \label{fig:median_unit_test_passes}
    \end{minipage}
    \vspace{1em} 
    \begin{minipage}[t]{0.50\linewidth}
        \centering
        \vspace{0pt} 
        \includegraphics[width=\linewidth]{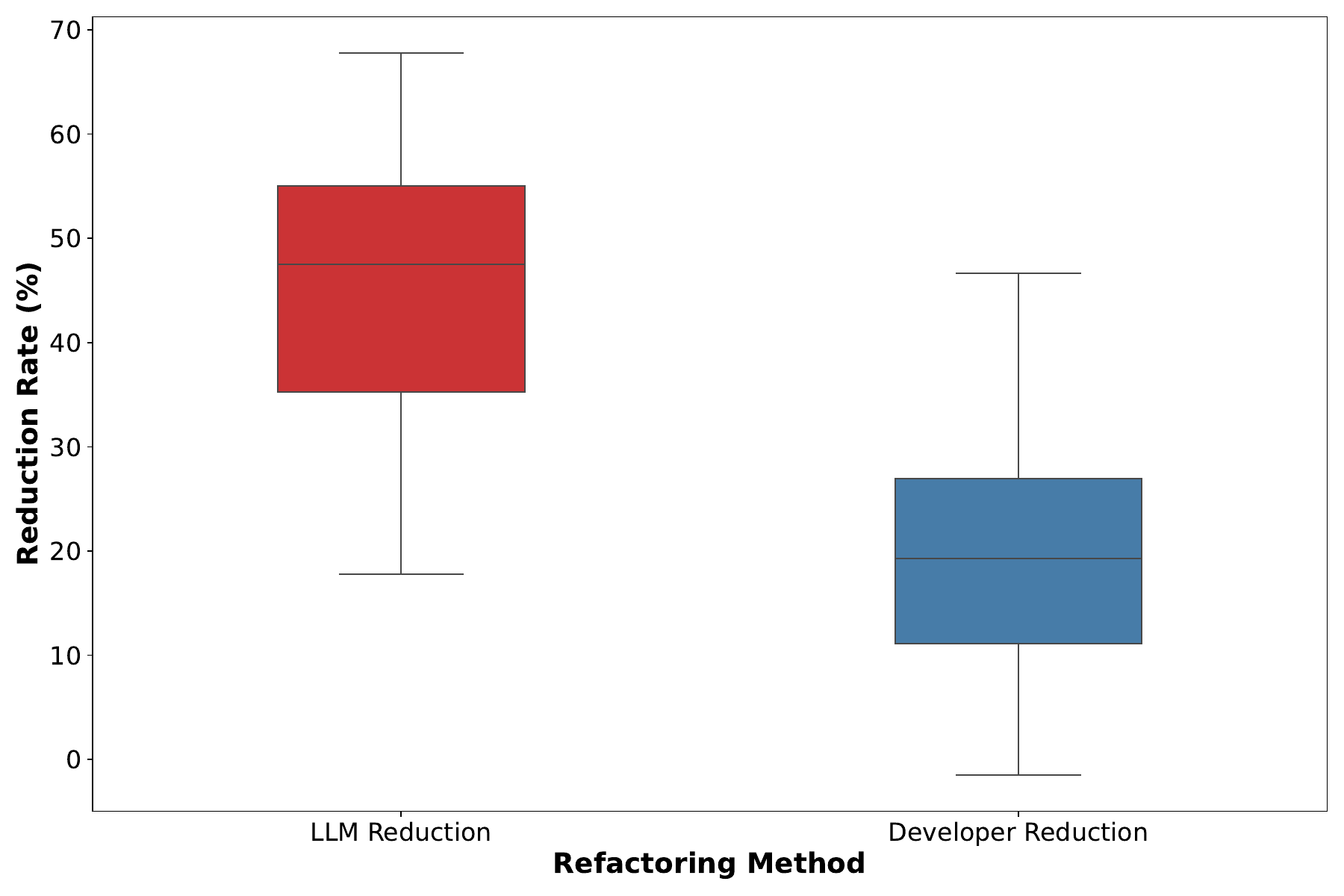}
        \caption{Distribution of Code Smell Reduction Rates Across 30 Projects for StarCoder2-Generated Refactorings and Developer Refactorings.}
        \label{fig:new_figure}
    \end{minipage}
\end{figure}
\textbf{Refactorings generated by StarCoder2 can sometimes propose changes that may affect the functionality of the code.} To provide a detailed comparison of the refactoring performance between StarCoder2 and developers, as shown in Table \ref{tab:unit_test_srr_comparison_starCoder2}, we present the unit test pass rates alongside the code smell reduction rate for pass@1, pass@3, and pass@5 scenarios. Developers achieve a 100\% unit test pass rate as all of their refactorings are accepted pull requests and maintain the same functionality of the code, while StarCoder2's performance varies across the pass metrics. StarCoder2 achieves a 28.36\% pass@1 for our experiments. We show the median unit test pass rates (\textit{i.e.}, pass@1, 3, 5) of the selected projects in Figure \ref{fig:median_unit_test_passes}. StarCoder2 refactored code has an average unit test pass rate of the 30 projects of \textbf{57.15\%} for the pass@5 metric. There is an average of 20.1\% improvement in unit test pass rate from pass@1 to pass@3 and an average of 8.7\% improvement in unit test pass rate from pass@3 to pass@5. The refactored code generated by StarCoder2 achieves a 57.15\% unit test pass rate at Pass@5, indicating substantial improvement over Pass@1 (28.36\%) and Pass@3 (8.7\%). \textbf{Therefore, it is recommended to have comprehensive testing when utilizing LLMs like StarCoder2 to ensure the correctness and integrity of the refactored code.} 

\textbf{StarCoder2 has better performance in code refactoring than developers when the generated code can pass the unit test.} Figure \ref{fig:smells_agg} shows the distribution of the total code smells before and after the application of developer refactoring and LLM refactoring in the 30 projects. The initial dataset representing unrefactored code contains 17,429 code smells, indicating several potential areas for improvement. Upon analyzing developer-driven refactoring, we observe 13,199 code smells, a code smell reduction rate of \textbf{24.27\%}. For code that StarCoder2 passes unit tests for, there is an initial 12,213 code smells before refactoring; we do not include code smells to calculate code smell reduction from code that StarCoder2 cannot successfully generate refactorings for (\textit{i.e.}, the refactorings pass unit tests). StarCoder2 reduces the initial number of code smells from 12,213 to 6,795 showing a code smell reduction rate of \textbf{44.36\%} for 30 projects. We show the distribution of code smell reduction rates across the 30 projects for both StarCoder2 and developers in Figure~\ref{fig:new_figure}. StarCoder2 reduces code smells at a 20.1\% higher rate than developers in our experiments. The results of the U-test for code smell reduction show a p-value of 0.003, indicating a significant difference in the distribution of code smells per commit before refactoring, after developer refactoring, and after LLM refactoring. StarCoder2 reduces code smells by 44.36\%, outperforming developers who achieve a 24.27\% reduction, making StarCoder2 20.1\% more effective in improving code quality through smell reduction.

Upon analyzing the 4.91\% increase in code smell reduction rate from Pass@1 to Pass@5, we observed a notable rise in the reduction rates of the \textit{Rebellious Hierarchy Smell} (5.18\%), along with improvements in reducing \textit{Long Method} (8.76\%) and \textit{Long Statement Smells} (6.32\%). The reduction rates for all other code smells remained consistent across Pass@1, Pass@3, and Pass@5. \textbf{This suggests that regenerating refactorings multiple times can enhance the effectiveness of code smell reduction, as the LLM may hallucinate for some refactoring generations that attempt to address complex code smells.}

\begin{table}
\centering
\caption{Comparison of Improvement in Metrics (\%) Over 30 Projects Between StarCoder2 and Developers. Cliff's Delta is shown for the Code Quality Metrics that are Significantly Improved by either StarCoder2 or Developers}
\label{tab:comparison}
\resizebox{\linewidth}{!}{%
\begin{tblr}{
  hline{1,14} = {-}{0.08em},
  hline{2,13} = {-}{0.05em},
}
\textbf{Metric}               & \textbf{Quality Attribute} & \textbf{LLM (Avg)} & \textbf{LLM (Median)} & \textbf{Dev (Avg)} & \textbf{Dev (Median)} & \textbf{Cliff's Delta (Interpretation)} \\
CountClassCoupled             & Coupling                   & 21.4               & 20.1                  & \textbf{24.1}      & 23.5                  & -                                    \\
CountClassCoupledModified     & Coupling                   & \textbf{18.3}      & 17.5                  & 16.8               & 16.2                  & -                                    \\
CountClassDerived             & Modularity                 & \textbf{17.9}      & 16.8                  & 15.2               & 14.7                  & -                                    \\
CountDeclClassVariable        & Modularity                 & 12.5               & 11.8                  & \textbf{14.9}      & 14.4                  & -                                    \\
CountDeclInstanceVariable     & Modularity                 & \textbf{19.7}      & 18.9                  & 17.0               & 16.5                  & -                                    \\
PercentLackOfCohesion         & Cohesion                   & \textbf{22.8}      & 21.7                  & 20.4               & 19.9                  & -                                    \\
PercentLackOfCohesionModified & Cohesion                   & \textbf{23.9}      & 22.8                  & 21.5               & 20.7                  & 0.42 (Medium)                        \\
AvgCyclomatic                 & Complexity                 & \textbf{17.4}      & 16.3                  & 14.6               & 14.0                  & 0.45 (Medium)                        \\
Cyclomatic                    & Complexity                 & \textbf{16.5}      & 15.4                  & 13.5               & 13.0                  & -                                    \\
MaxCyclomatic                 & Complexity                 & \textbf{15.8}      & 14.9                  & 13.3               & 12.8                  & -                                    \\
SumCyclomatic                 & Complexity                 & \textbf{18.6}      & 17.4                  & 15.9               & 15.1                  & 0.47 (Medium)                        \\
\textbf{Average}              & -                          & \textbf{19.32}     & 18.2                  & 17.46              & 16.9                  & -                                    
\end{tblr}
}
\end{table}
\textbf{StarCoder2 excels in improving key code metrics related to cohesion, modularity, and complexity, often surpassing developers in these areas.} The results presented in Table \ref{tab:comparison} provide a quantitative analysis of the refactoring capabilities of StarCoder2 compared to developers, where a higher percentage reduction indicates a greater ability to improve code quality. StarCoder2 achieves a \textbf{19.7\%} average reduction in instance variables (\textit{i.e.}, CountDeclInstanceVariable), outperforming developers, who achieve a \textbf{17.0\%} reduction. Similarly, StarCoder2 leads in reducing PercentLackOfCohesion by \textbf{22.8\%} compared to \textbf{20.4\%} for developers. These results suggest that StarCoder2 is particularly effective at reducing code elements that contribute to poor cohesion, leading to more modular and well-structured code. StarCoder2 shows a significant improvement in reducing PercentLackOfCohesionModified, with a \textbf{23.9\%} reduction compared to \textbf{21.5\%} for developers, demonstrating a moderate effect size of \textbf{0.42} (Cliff’s Delta). This indicates that StarCoder2 consistently improves cohesion after modifications, more so than developers. StarCoder2 excels in reducing cohesion-related code metrics, such as instance variables and cohesion percentages, outperforming developers in these areas. It shows a 19.7\% reduction in instance variables and a 22.8\% reduction in cohesion issues, indicating better code modularity and structure.
However, developers outperform StarCoder2 in reducing the number of coupled classes (\textit{i.e.}, CountClassCoupled and CountDeclClassVariable), with average reductions across 30 projects of \textbf{24.1\%} and \textbf{14.9\%}, respectively. This suggests that \textbf{while StarCoder2 excels in several areas, developers may still have an edge in refactoring tasks that require a deep understanding of class dependencies and variable declarations.}

\textbf{When examining cyclomatic complexity, StarCoder2 once again outperforms developers.} The model reduces AvgCyclomatic by \textbf{17.4\%} and SumCyclomatic by \textbf{18.6\%}, compared to reductions of \textbf{14.6\%} and \textbf{15.9\%} by developers, respectively. Notably, the reduction in SumCyclomatic has a medium effect size of \textbf{0.47}, highlighting StarCoder2’s effectiveness in simplifying code logic across entire projects. The model's ability to reduce complexity on both individual methods and overall project levels suggests that it may contribute to easier maintainability over time. We reject the null hypothesis H\textsubscript{0} as StarCoder2 performs significantly better in code smell reduction and various code quality metric improvements compared with developers.

We use the best refactoring generation from pass@5 to answer the remaining research questions. Pass@5 shows better performance than pass@1 and pass@3 in passing unit tests and therefore results in more valid refactorings to analyze. Taking the best refactoring generation from pass@5 results in more data for analyzing refactoring types performed by the LLM as well as code smell types reduced by the LLM.

\begin{tcolorbox}[colback=blue!5!white, colframe=blue!75!black, title=Summary of Results]
StarCoder2 outperforms developers across most evaluated code metrics, achieving an average reduction across all metrics of \textbf{19.32\%} compared to \textbf{17.46\%} for developers, demonstrating its strength in automating complex refactoring tasks. StarCoder2 improves unit test pass rates significantly, reaching \textbf{57.15\%} at Pass@5, though developers maintain a \textbf{100\%} pass rate. StarCoder2 excels in code smell reduction, reducing smells by \textbf{44.36\%}, which is \textbf{20.1\%} higher than developers, and particularly effective in addressing smells like long methods and rebellious hierarchy. It also shows superior performance in reducing cohesion and complexity metrics, achieving greater modularity and structure. However, developers outperform StarCoder2 in reducing class coupling, highlighting their advantage in tasks requiring a deeper understanding of class dependencies. Despite this, StarCoder2's ability to simplify code logic and reduce cyclomatic complexity makes it a valuable tool for improving code maintainability.
\end{tcolorbox}

\subsection{\textbf{RQ2. Which types of code smells are most effectively reduced by LLMs or developers?}}
\subsubsection{\textbf{Motivation}}
Quality refactorings are linked with the elimination of code smells~\cite{10.1145/2950290.2950305, 10.1145/3106237.3106259, tsantalis2020refactoringminer}. These smells can range from simple syntactic issues, such as \textit{Long Statement} and \textit{Empty Catch Clause}, to more complex design flaws, such as \textit{Insufficient Modularization} and \textit{Multifaceted Abstraction}. Understanding which types of code smells are most effectively reduced by developers or by LLMs is crucial for identifying their strengths and limitations. Developers have domain knowledge and an understanding of software design principles, while LLMs can apply their extensive training on code patterns and idioms to address repetitive and rule-based smells. Given that different types of code smells vary in complexity and context dependency, this research question aims to explore how well LLMs and developers perform across various categories of code smells. By investigating which code smells each refactoring approach excels at reducing, we can better understand the strengths of the LLM versus human developers in code refactoring.

\subsubsection{\textbf{Approach}}
To analyze the effectiveness of refactorings performed by developers and StarCoder2 in reducing different types of code smells, we compare the performance of StarCoder2 and developers in code smell reductions and quantify the differences for each specific type of code smell. Specifically, we examine each commit that contains at least one instance of a code smell, ensuring meaningful comparisons. The detailed steps conducted are provided below:

\textbf{Comparison of differences: }
Since our data does not follow a normal distribution, we use the Mann-Whitney U-test~\cite{doi:https://doi.org/10.1002/9780470479216.corpsy0524} to examine if the distributions of reductions between StarCoder2 and developers originate from different distributions. The U-test evaluates whether one distribution tends to have higher values than the other, based on the rank of each data point. Each data point in the distribution represents the reduction of a specific code smell within a single project (\textit{i.e.}, the reduction rate after refactoring by either StarCoder2 or developers for that code smell within that project). Furthermore, we quantify the differences in the effect size using Cliff's delta (\(\delta\))~\cite{meissel2024cliffs} as discussed in Section~\ref{sec:approach_rq1}.
 
\subsubsection{\textbf{Findings}}

\begin{table}
\centering
\caption{Comparison of Code Smell Reductions Between StarCoder2 and Developers.}
\label{tab:code_smell_reductions}
\resizebox{\linewidth}{!}{%
\begin{tblr}{
  vline{2} = {-}{},
  hline{1-2,10,18} = {-}{},
}
\textbf{Code Smell Category} & \textbf{Code Smell} & \textbf{Better Reduction} & \textbf{Cliff's Delta ($\delta$)} & \textbf{Interpretation}\\
Design & Unutilized Abstraction & LLM & 0.5366 & Large\\
 & Unnecessary Abstraction & LLM & 0.5130 & Large\\
 & Broken Hierarchy & LLM & 0.5608 & Large\\
 & Cyclic-Dependent Modularization & LLM & 0.4815 & Large\\
 & Broken Modularization & Developer & 0.6417 & Large\\
 & Deficient Encapsulation & Developer & 0.5587 & Large\\
 & Multifaceted Abstraction & Developer & 0.4636 & Medium\\
 & Insufficient Modularization & Developer & 0.6768 & Large\\
Implementation & Long Statement & LLM & 0.5406 & Large\\
 & Magic Number & LLM & 0.6310 & Large\\
 & Empty Catch Clause & LLM & 0.6018 & Large\\
 & Complex Conditional & LLM & 0.5355 & Large\\
 & Long Parameter List & LLM & 0.5703 & Large\\
 & Long Identifier & LLM & 0.5669 & Large\\
 & Complex Method & LLM & 0.5271 & Large\\
 & Missing Default & Developer & 0.4488 & Medium
\end{tblr}
}
\end{table}

\textbf{The LLM outperforms developers in reducing 10 of the 16 types of code smells.}As illustrated in Table \ref{tab:code_smell_reductions}, StarCoder2 can reduce code smells in 7 out of 8 types of code smells in the implementation category with a significantly large effect size comparing the reduction of code smells. In particular, StarCoder2 excels in addressing syntactic and pattern-based smells, such as \textit{Long Statement}, \textit{Long Parameter List}, \textit{Long Identifier}, \textit{Empty Catch Clause}, and \textit{Magic Number}, which follow more regular idiomatic, and repetitive patterns. An explanation of each of the code smells found in the table is provided~\cite{A_taxonomy_of_software_smells}. This finding suggests that \textbf{StarCoder2 is particularly effective at addressing issues related to implementation, where structured, rule-based corrections can be applied.}  

\textbf{Developers show a better performance in reducing complex and context-sensitive code smells, particularly those related to modularization.} As shown in Table \ref{tab:code_smell_reductions}, developers outperform the LLM in reducing \textit{Missing Default}, \textit{Insufficient Modularization}, and \textit{Multifaceted Abstraction}, which require a deeper understanding of the overall software architecture and implementation principles. The large effect sizes, as indicated by the Cliff’s Delta (\(\delta\)) values, for these code smells show that \textbf{developers have a significantly large advantage for handling complex design code smells.} \textbf{In the design category, StarCoder2 outperforms developers in 4 out of 8 design-related code smells (\textit{i.e.}, \textit{Unutilized Abstraction}, \textit{Unnecessary Abstraction}, \textit{Broken Hierarchy}, and \textit{Cyclic-Dependent Modularization})}. The LLM's effectiveness in reducing design-related smells suggests that it can eliminate unnecessary complexity and optimize code structure when the issues are relatively straightforward. However, developers outperform StarCoder2 in addressing more intricate design problems like \textit{Deficient Encapsulation} and \textit{Broken Modularization}, which require deeper reasoning about class dependencies, encapsulation, and how components interact at a design level.

\begin{tcolorbox}[colback=blue!5!white, colframe=blue!75!black, title=Summary of Results]
Our findings suggest that while StarCoder2 performs better at handling implementation-level code smells and some straightforward design issues, such as \textit{Long Statement, Magic Number, Long Identifier}, etc., developers are better at handling more complex, context-dependent code smells, particularly those related to modularization and encapsulation. This points to the potential value of leveraging the strengths of both LLMs and developers for comprehensive code smell reduction and to improve overall software quality.
\end{tcolorbox}

\subsection{\textbf{RQ3. Which refactoring types are most effective by LLMs or by developers for improving code quality?}}
\subsubsection{\textbf{Motivation}}
Refactoring types refer to the specific categories of code modifications aimed at improving the structure, readability, and maintainability of software without altering its external behavior. Examples include renaming variables for clarity, extracting methods to reduce code duplication, or reorganizing classes to better align with design principles. In this research question, we aim to understand the differences in the application of various refactoring types as quality improvements are achieved by both developers and StarCoder2. This comparison enables us to determine whether StarCoder2 or developers focus on different refactoring types and highlights their respective strengths in the application of different refactoring types. This information can guide which refactoring techniques should be prioritized in automated tools and development practices. 

\subsubsection{\textbf{Approach}}
To analyze the different types of refactorings applied and assess their impact on code quality, we first collect the refactorings performed by StarCoder2 and developers. We then analyze the statistical differences in refactoring types and their effects on code quality between StarCoder2 and developers. The steps of this analysis are detailed below.

\textbf{Quality and Impact Measurement:} We use RMiner 3.0~\cite{tsantalis2020refactoringminer} to extract all the refactorings applied by both StarCoder2 and developers across all projects. RMiner is the most comprehensive refactoring detection tool for Java, capable of identifying up to 102 refactoring types, with a precision of 99.8\% and a recall of 98.1\%~\cite{Tsantalis:ICSE:2018:RefactoringMiner, tsantalis2020refactoringminer}.
Then, we examine the distribution of frequency of each refactoring type performed across all projects together to understand any differences in applying refactoring strategies between the LLM and developers. Next, we analyze whether StarCoder2 and developers perform and prioritize different refactoring types. By calculating the code smell reducrtion rate, we evaluate the effectiveness of the refactorings performed by developers and StarCoder2 in targeting specific types of smells. This analysis reveals whether LLM-driven refactorings and developer-driven refactorings focus on reducing the same types of code smells or whether their strategies differ in terms of addressing different issues within the code.
Moreover, we assess how refactorings performed by StarCoder2 and developers impact complexity, cohesion, coupling, and modularity. By comparing the changes in these metrics before and after refactoring, we aim to determine whether StarCoder2 and developers emphasize different aspects of code quality improvement such as reducing complexity or improving modularity. This analysis helps identify whether LLMs prioritize improvements in certain areas of code quality more than developers, or vice versa.

\textbf{Significance and Size Difference Measurement:}
To evaluate whether the differences in refactoring types performed between StarCoder2 and developers are statistically significant, we perform a Mann-Whitney U-test~\cite{doi:https://doi.org/10.1002/9780470479216.corpsy0524}. Each data point in our analysis represents either the reduction in a specific code smell for each refactoring type after refactoring by StarCoder2 or by developers. Additionally, we analyze improvements in code quality metrics following refactoring by both StarCoder2 and developers. We first use the U-test to check whether the distributions of code smell reduction between StarCoder2 and developers are statistically different. If the U-test reveals a significant difference (p-value < 0.05), we conclude that refactorings performed by StarCoder2 and the developer significantly target different areas of the code, indicated by the different types of code smells. To further quantify the extent of this difference, we apply Cliff’s delta (\(\delta\))~\cite{meissel2024cliffs} to measure the effect size. We discuss small, medium, and large effect sizes in Section \ref{sec:cliff's delta}. A large Cliff’s delta value indicates that StarCoder2 and developers exhibit substantial differences in their code smell reduction, while a smaller value suggests a more subtle difference between the two approaches. We take the same approach when evaluating the impact of StarCoder2 or developer refactoring on code quality metric improvement.

\subsubsection{\textbf{Findings}}
\label{sec:rq3_findings}

\begin{figure*}[htbp]
    \centering
    \includegraphics[width=\linewidth]{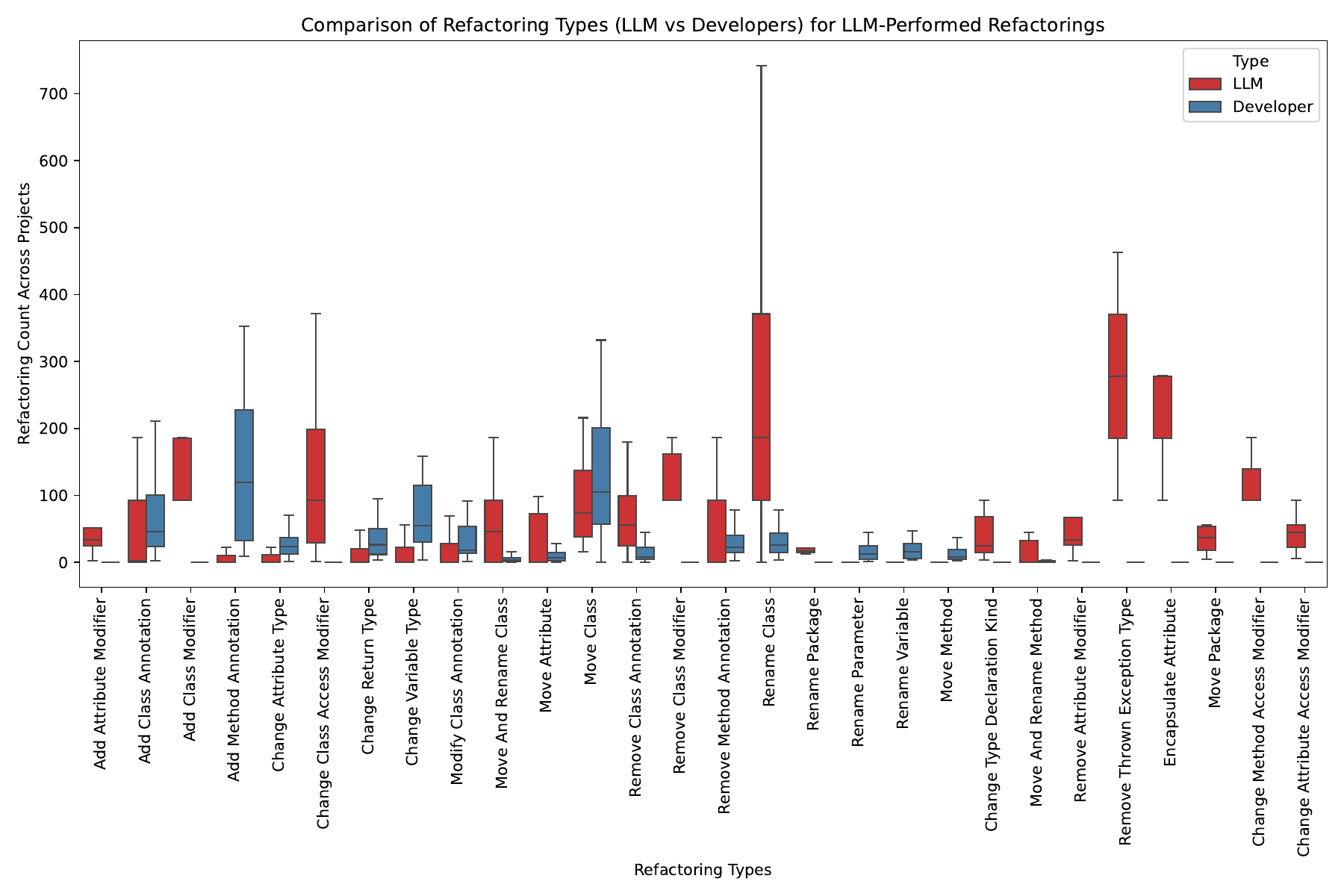}
    \caption{Distribution of Frequencies of all Refactoring Types Used by StarCoder2 Across 30 Projects.}
    \label{fig:refactoring_types_boxplot}
\end{figure*}

\textbf{There are distinct preferences in performing various refactoring types between StarCoder2 and developers.} Figure \ref{fig:refactoring_types_boxplot} shows the distribution of all refactoring types StarCoder2 performs. StarCoder2 demonstrates a higher frequency of certain refactorings which can be detected by following syntactic rules, such as \textit{Rename Method} and \textit{Extract Method}, compared to developers. Conversely, developers more frequently perform refactorings that require more checks on dependencies and cause a larger impact on code change propagation, such as Move Method and Change Attribute Access Modifier.

\begin{table}[t]
    \centering
    \caption{Comparison of Different Refactoring Type's Effect on Code Smell Reduction Between StarCoder2 and Developers}
    \setlength{\tabcolsep}{4pt} 
    \scriptsize
    \begin{tabular}{lccc}
        \toprule
        \textbf{Refactoring Type} & \textbf{Better Reduction} & \textbf{Cliff's Delta (\(\delta\))} & \textbf{Interpretation} \\
        \midrule
        Remove Class Annotation & LLM & 0.6969 & Large \\
        Move Class & LLM & 0.6060 & Large \\
        Add Method Annotation & LLM & 0.9117 & Large \\
        Remove Method Annotation & LLM & 0.9619 & Large \\
        Rename Class & LLM & 0.5871 & Large \\
        Move Attribute & Developer & 0.7832 & Large \\
        Modify Class Annotation & LLM & 0.9589 & Large \\
        Add Class Annotation & LLM & 0.8539 & Large \\
        Move And Rename Method & LLM & 0.9394 & Large \\
        Change Attribute Type & Developer & 0.7796 & Large \\
        Move And Rename Class & LLM & 0.5048 & Large \\
        \bottomrule
    \end{tabular}
    \label{tab:refactoring_types}
\end{table}

Table~\ref{tab:refactoring_types} presents a comparison of refactoring types performed by either StarCoder2 or developers that significantly reduce code smells more than the other (p-value \textless 0.05). The data shows that StarCoder2 performs significantly better in reducing code smells for refactoring types such as Remove Class Annotation, Remove Method Annotation, and Rename Class. The Cliff's Delta (\(\delta\)) values for these types indicate a large effect size. StarCoder2 is particularly effective at eliminating annotations and renaming classes in ways that reduce code smells more efficiently than human developers. This is evidenced by the large effect size for Move Attribute and Change Attribute Type. The developers perform better at the attribute level which involves data dependencies; these refactorings often require a deeper understanding of the code's context and architecture. \textbf{Developers tend to engage in more types of refactoring that involve dependency checks and have a greater impact on the code, demonstrating superior performance in managing complex structural changes.}

\begin{table}[t]
    \centering
    \caption{Comparison of Different Refactoring Types' Effect on Code Metrics Improvement}
    \setlength{\tabcolsep}{4pt} 
    \scriptsize
    \begin{tabular}{lccc}
        \toprule
        \textbf{Refactoring Type} & \textbf{Better Improvement} & \textbf{Cliff's Delta (\(\delta\))} & \textbf{Interpretation} \\
        \midrule
        Rename Class & LLM & 0.6123 & Large \\
        Add Class Annotation & LLM & 0.8517 & Large \\
        Encapsulate Attribute & Developer & 0.6798 & Large \\
        Change Return Type & LLM & 0.8032 & Large \\
        Change Attribute Access Modifier & Developer & 0.7215 & Large \\
        \bottomrule
    \end{tabular}
    \label{tab:code_metrics_improvement}
\end{table}

To highlight the differences in refactoring patterns between StarCoder2 and developers, we categorized the refactorings and observed distinct focuses. 
\begin{itemize}
    \item  \textbf{Access control refactorings, which are changes to access modifiers and encapsulation of attributes, are unique to StarCoder2, which frequently modifies access levels and encapsulates attributes, addressing visibility and encapsulation issues in a rule-based manner.} For instance, StarCoder2 performs refactorings like \textit{Change Method Access Modifier} and \textit{Encapsulate Attribute}, focusing on improving data hiding and access control.
    \item\textbf{Developers handle more complex method and attribute refactorings such as \textit{Extract Method} and \textit{Inline Variable}, which involve restructuring code for better modularity and readability.} These types of refactorings require a deeper understanding of code context. Class-level refactorings also show divergence, with StarCoder2 applying simpler changes like \textit{Change Type Declaration Kind} and \textit{Add Class Modifier}, while developers performed more intricate refactorings like \textit{Extract Superclass} and \textit{Pull Up Method}, requiring a deeper understanding of class hierarchies. Additionally, package and file movement refactorings are handled differently: StarCoder2 focuses on high-level tasks such as \textit{Move Package}, while developers engage in more granular changes such as \textit{Move Source Folder}. 
\end{itemize}
    The full details of all refactorings unique to StarCoder2 and unique to developers are included in the replication package

\begin{tcolorbox}[colback=blue!5!white, colframe=blue!75!black, title=Summary of Results]
StarCoder2 excels in automated, syntactic refactoring types, while developers focus on more complex, structural changes, demonstrating complementary strengths between the two approaches. A combined approach leveraging both LLMs and developers could potentially offer the most comprehensive strategy for code refactoring.
\end{tcolorbox}

\subsection{\textbf{RQ4. How does prompt engineering affect the quality of LLM-generated refactorings?}}
\subsubsection{\textbf{Motivation}}
Our results from the third research question show that developers often apply more refactoring types that reduce code smells and improve code quality, but LLMs perform fewer refactoring types than developers. By using prompt engineering~\cite{10.1145/3544548.3581388}, which involves designing the input prompts to guide the LLM's responses more effectively, we aim to enhance the LLM’s ability to perform more complex and a wider variety of refactoring types—specifically for refactoring types that the LLM either do not perform or perform less frequently compared with developers. The objective is to determine whether modifying the instructions or providing additional context in the prompt can encourage StarCoder2 to execute more complex refactoring types, typically performed by developers, thus improving the code quality.

\subsubsection{\textbf{Approach}}
To enhance StarCoder2's refactoring experience, we aim to evaluate different prompting techniques and identify the most effective ones for improving code quality. In the following, we describe our approach.

\textbf{Prompt strategy selection:}
We begin by selecting prompting strategies that vary in their level of detail, guidance, and contextual information. 
We first identify refactoring types that are consistently performed by developers but are missing in StarCoder2's output as the results from RQ3 (Section \ref{sec:rq3_findings}). Once we identify these refactoring types, we design specific prompts to guide StarCoder2 in applying them. These prompts are categorized as follows:
\begin{itemize}
    \item Zero-shot Prompt (baseline): This prompt is the baseline used in RQ1. We compare the performance of the following prompts with this baseline shown in Figure~\ref{fig:model_prompt}. 
    \item Chain-of-thought: Shown in Figure~\ref{fig:model_prompt_cot}, this prompt provides suggestions for refactorings, along with the explanation of that refactoring type, by including common refactoring patterns developers perform that the LLM does not. For example, the prompt may instruct, "Consider using Extract Method or Inline Method," guiding StarCoder2 toward specific types of refactorings that developers perform with more efficiency.
   
    \item One-Shot Prompt: Shown in Figure~\ref{fig:model_prompt_one}, in this approach we provide an example of a correctly refactored code snippet by the developer before asking StarCoder2 to refactor a new piece of code. We provide the LLM with an example by the developer that uses the same refactoring type on the commit the LLM will refactor. This technique guides the model to follow the same approach as in the provided example. For instance, the prompt would present a refactored code snippet where a method has been successfully extracted and then ask the model to perform similar refactorings on a new code sample. This prompt aims to teach the model by example, increasing the likelihood of correct and consistent refactoring operations.
    \begin{figure}[h]
        \centering
        \begin{minipage}[b]{0.45\linewidth} 
            \centering
            \fbox{
                \parbox{0.95\textwidth}{
                    \scriptsize
                    \textbf{Instruction} \\
                    You are a powerful model specialized in refactoring Java code. Code refactoring is
                    the process of improving the internal structure, readability, and maintainability of a software codebase without altering its external behavior or functionality. You must output a refactored version of the code. Explain the steps you took to refactor the code and why you selected the refactoring type/types you did.
    
                    \vspace{0.3cm}
                    \textbf{Prompt} \\
                    \textbf{\# Suggested refactoring types:} \\
                    \{refactorings developers performed on this commit with definitions\} \\
                    \textbf{\# unrefactored code snippet(java):} \\
                    \{code\_segment\_before\_refactoring\}
                    
                    \vspace{0.3cm}
                    \textbf{\# refactored version of the same code snippet:}
                    
                }
            }
            \caption{Chain-of-Thought Prompt Used to Instruct StarCoder2 to Conduct Refactoring}
            \label{fig:model_prompt_cot}
        \end{minipage}
        \hspace{0.05\linewidth} 
        \begin{minipage}[b]{0.45\linewidth} 
            \centering
            \fbox{
                \parbox{0.95\textwidth}{
                    \scriptsize 
                    \textbf{Instruction} \\
                    You are a powerful model specialized in refactoring Java code. Code refactoring is
                    the process of improving the internal structure, readability, and maintainability of a software codebase without altering its external behavior or functionality. You must output a refactored version of the code.
    
                    \vspace{0.3cm}
                    \textbf{Prompt} \\
                    \textbf{\# unrefactored code snippet(java):} \\
                    \{code\_segment\_before\_refactoring\} \\
                    \textbf{\# refactored version of the same code snippet:} \\
                    \{developer\_code\_after\_refactoring\}
                    
                    \vspace{0.3cm}
                    \textbf{\# unrefactored code snippet(java):} \\
                    \{code\_segment\_before\_refactoring\}

                    \vspace{0.3cm}
                    \textbf{\# refactored version of the same code snippet:}
                }
            }
            \caption{One-shot Prompt Used to Instruct StarCoder2 to Conduct Refactoring}
            \label{fig:model_prompt_one}
        \end{minipage}
    \end{figure}
\end{itemize}

\textbf{Quality impact evaluation:}
To evaluate the impact of different prompting techniques, we compare the quality of refactorings generated by StarCoder2 across the unit test pass rate which measures functional correctness and compilability, the code smell reduction rate, and code quality metrics. The comparison allows us to assess whether certain prompts improve the model’s performance on refactorings that developers typically perform better at. For each commit in our dataset, we generate refactorings using the 2 categories of prompts. We evaluate the effectiveness of these prompts by comparing their ability to improve code quality metrics and reduce code smells, with particular attention to whether StarCoder2 begins to perform refactorings it previously did not perform.
To assess whether the prompting techniques lead to statistically significant improvements, we perform the Scott-Knott test~\cite{scott-knott} across the three distributions: zero-shot, one-shot, and chain-of-thought. The Scott-Knott~\cite{scott-knott} test is a hierarchical clustering method that partitions distributions into statistically distinct groups, helping to identify whether the differences between groups are meaningful~\cite{scott-knott}. The data samples consist of quality improvements (\textit{i.e.}, code smell reduction rate, and percentage of improvement in code metrics) achieved by StarCoder2 under different prompt conditions for each project. We compare the zero-shot prompt, one-shot prompt, and chain-of-thought prompt for their performance in each code metric and code smell reduction to determine if the prompting techniques affect code refactoring performance and compare the performance of three prompting techniques.

\subsubsection{\textbf{Findings}}
\textbf{Applying one-shot prompting and chain-of-thought have more significant influences on code smell reduction than applying zero-shot prompting.} Table \ref{tab:prompt_comparison} summarizes the results for the unit test pass rates and code smell reduction rates across the 3 prompting methods: zero-shot, one-shot, and chain-of-thought prompts. Table \ref{tab:prompt_comparison_metrics} shows the results for code quality metric improvement rates across the 3 prompting methods.

\begin{table} [t]
\scriptsize
\centering
\caption{Scott-Knott Test Comparing Unit Test Pass Rate and Code Smell Reduction Rates (SRR) across Different Prompting Techniques}
\label{tab:prompt_comparison}
\begin{tblr}{
  column{even} = {c},
  column{3} = {c},
  hline{1,5} = {-}{0.08em},
  hline{2} = {-}{0.05em},
}
\textbf{Prompting Method} & \textbf{Unit Test Pass Rate (\%)} & \textbf{SRR (\%)} & \textbf{Scott-Knott Rank (SRR)}\\
Zero-shot & 28.36\% & 39.45\% & 2\\
Chain-of-thought & 32.22\% & 42.34\% & 1\\
One-shot & \textbf{34.51\%} & 42.97\% & 1
\end{tblr}
\end{table}

\begin{table}
\centering
\scriptsize
\caption{Scott-Knott Test Comparing Average Improvement in Each Code Metrics (\%) across Different Prompting Methods}
\label{tab:prompt_comparison_metrics}
\resizebox{\linewidth}{!}{%
\begin{tblr}{
  column{even} = {c},
  column{3} = {c},
  column{5} = {c},
  column{7} = {c},
  hline{1,14} = {-}{0.08em},
  hline{2,13} = {-}{0.05em},
}
\textbf{Metric} & \textbf{Zero-shot} & \textbf{Scott-Knott Rank (Zero-shot)} & \textbf{Chain-of-thought} & \textbf{Scott-Knott Rank (CoT)} & \textbf{One-shot} & \textbf{Scott-Knott Rank (One-shot)}\\
CountClassCoupled & 21.4 & 2 & \textbf{23.4} & 1 & 22.7 & 1\\
CountClassCoupledModified & 18.3 & 3 & 19.2 & 2 & \textbf{20.1} & 1\\
CountClassDerived & 17.9 & 2 & 18.1 & 2 & \textbf{18.3} & 2\\
CountDeclClassVariable & 12.5 & 2 & \textbf{14.0} & 1 & 13.8 & 1\\
CountDeclInstanceVariable & 19.7 & 2 & \textbf{20.2} & 2 & 19.8 & 2\\
PercentLackOfCohesion & 22.8 & 2 & \textbf{23.5} & 1 & 23.1 & 2\\
PercentLackOfCohesionModified & 23.9 & 2 & 24.1 & 2 & \textbf{24.2} & 2\\
AvgCyclomatic & 17.4 & 2 & 18.0 & 1 & \textbf{18.4} & 1\\
Cyclomatic & 16.5 & 2 & 17.0 & 1 & \textbf{17.3} & 1\\
MaxCyclomatic & 15.8 & 2 & 16.3 & 1 & \textbf{16.6} & 1\\
SumCyclomatic & 18.6 & 2 & 19.1 & 1 & \textbf{19.3} & 1\\
\textbf{Average} & 19.32 & 2 & 19.99 & 1 & \textbf{20.15} & 1
\end{tblr}
}
\end{table}

\begin{itemize}
    \item \textbf{One-shot prompting}: The one-shot prompt method achieves the highest performance, with a unit test pass rate of \textbf{34.51\%} and a code smell reduction rate of \textbf{42.97\%} (Scott-Knott Rank 1). One-shot prompting is particularly effective at reducing code complexity metrics such as \textit{AvgCyclomatic}, \textit{Cyclomatic}, and \textit{MaxCyclomatic}, all of which achieve the highest Scott-Knott rank (Rank 1), indicating consistent improvements. Additionally, this method yields a significant reduction in \textit{CountClassCoupledModified} (Rank 1), contributing to better modularity.

    \item \textbf{Chain-of-thought prompting}: This method also performs well, with a unit test pass rate of \textbf{32.22\%} and a code smell reduction rate of \textbf{42.34\%} (Scott-Knott Rank 1). It demonstrates notable improvements in metrics like \textit{CountClassCoupled} (Rank 1) and \textit{CountDeclClassVariable} (Rank 1), with statistically significant differences. Furthermore, chain-of-thought prompting increases the variety of refactoring types performed by the LLM, adding seven new refactoring types not observed with zero-shot prompting: \textit{Extract Method, Rename Method, Extract Variable, Inline Method, Add Parameter, Extract Class, and Parameterize Variable}. This method enhances the model's ability to generate a broader range of refactorings by providing specific instructions and definitions.

    \item \textbf{Zero-shot prompting}: Exhibits the lowest performance, with a unit test pass rate of \textbf{28.36\%} and a code smell reduction rate of \textbf{39.45\%} (Scott-Knott Rank 2). This suggests that StarCoder2's ability to autonomously generate high-quality refactorings is limited when given vague or minimal instructions. Metrics such as \textit{AvgCyclomatic}, \textit{Cyclomatic}, and \textit{MaxCyclomatic} consistently receive a lower rank (Rank 2), indicating less significant improvements compared to one-shot and chain-of-thought prompting.

\end{itemize}

One-shot prompting consistently yields the best performance across various metrics, including code complexity and modularity, by providing a concrete example of refactoring. Chain-of-thought prompting follows closely, especially in terms of expanding the types of refactorings performed and showing improvements in coupling and class variables. Both methods significantly outperform the zero-shot baseline, reinforcing the importance of prompt design in optimizing the performance of LLMs for refactoring tasks.

\begin{tcolorbox}[colback=blue!5!white, colframe=blue!75!black, title=Summary of Results]
Our findings highlight the importance of prompt design in leveraging the full potential of LLMs for code refactoring tasks. Incorporating examples (\textit{i.e.}, one-shot prompt) or providing more explicit instructions (\textit{i.e.}, chain-of-thought prompt) significantly improves both the functional correctness and overall quality of the refactorings generated by StarCoder2. The variation in code smell reduction rates and unit test pass rates across different prompting methods underscores the value of prompt engineering in optimizing LLM performance for complex tasks like refactoring.
\end{tcolorbox}
\section{Threats to Validity}
\label{sec:threads_to_validity}
In this section, we outline potential threats to the validity of our study and the measures taken to mitigate them.

\textbf{Threats to Internal Validity.} Specify the validity of the methods applied in our study. While the performance of StarCoder2 could theoretically be influenced by specific hardware and software configurations, we ensured that all experiments were conducted under identical conditions across the same server, using consistent hardware and software setups. Therefore, running the experiments on a different server should not significantly affect the results, as long as comparable computational resources are used. This mitigates the concern of hardware or server variability as a threat to internal validity.

A potential threat to internal validity is the occurrence of model hallucination, where StarCoder2 may generate code that appears correct but is not logically or syntactically valid. Such hallucinations could alter the accuracy of our results, particularly in evaluating unit test pass rates and refactoring effectiveness. We account for this by verifying the output through unit test evaluations and inspecting for invalid code, but hallucinations remain an inherent risk in LLM-generated code.

We utilize a randomly selected set of representative commits from our projects. However, our results may be influenced by the random selection of commits, as variations in the project settings or the choice of different subsets of commits could lead to differing outcomes. Some of the refactorings might span multiple commits, however, our approach focuses on analyzing refactorings within a single commit. This limitation could result in missing refactorings that are spread over several commits, potentially underestimating the complexity or scope of certain changes. Although we select projects not included in the StarCoder2 training dataset, there is still a possibility that similar code patterns or practices could be present in the training set of the model, which might give the LLM an advantage.

\textbf{Threats to External Validity.} Indicate the generalizability of our approach. Our experiments are conducted on a limited set of open-source Java projects from the Apache dataset. Therefore, the findings may not be directly applicable to projects in other programming languages or domains. We make the replication package publicly available to allow other researchers to test across a more diverse set of projects and languages.
Our study is based on the latest open-source version of StarCoder2-15B-Instruct-v0.1, as of the date of our study. While our results reflect the capabilities of this specific version, our evaluation approach is designed to be adaptable. It can be used to assess future versions of StarCoder2 as well as other LLMs from different vendors. Nevertheless, different LLMs or different versions of StarCoder may generate different results.

\textbf{Threat to Construct Validity.} Concern about our feature selection and our data preparation. We primarily rely on code smell reduction, code quality metrics improvement, and unit test pass rates as indicators of refactoring quality. However, these metrics may not fully capture all dimensions of code quality, such as readability and performance. This limitation may affect the robustness of our conclusions in a broader range of quality indicators. We used Rminer3.0, DesigniteJava, and Understand to gather our metrics for analysis which are the best in the state of practices for Java, however different tools may get different results.

\section{Related Work}
\label{sec:related_work}
The application of Large Language Models (LLMs) in software engineering, particularly in code refactoring, has garnered significant attention. In this section, we categorize the related work into three key areas: (1) code generation approaches before and after the rise of LLMs, (2) code refactoring using LLMs, and (3) prompt engineering and fine-tuning techniques to enhance LLM performance.

\subsection{Code Generation Approaches: Pre-LLM and Post-LLM Era}

Before the development of LLMs, code generation and repair primarily relied on rule-based systems and traditional program synthesis techniques. These approaches typically use formal methods and predefined rules to generate code, which limits their ability to handle complex programming tasks. For instance, DeepCoder~\cite{DBLP:journals/corr/BalogGBNT16} demonstrates an early machine learning-based approach to code generation, synthesizing programs from input-output pairs, but struggling with more complex scenarios. Similarly, AlphaCode~\cite{li2022competition} represents a more recent effort in competitive programming, where LLMs like AlphaCode have shown remarkable improvements in generating functional code from natural language descriptions. 

With the rise of LLMs, a new standard for code generation has emerged. Studies, such as Codex~\cite{10.1109/ICSE48619.2023.00128}, have demonstrated the potential of LLMs to generate code through natural language prompts. Codex, for example, has been evaluated for its ability to repair programs using test-based repair tools, showcasing strengths in code completion but also highlighting limitations in handling complex tasks due to a lack of semantic understanding. Xu et al.~\cite{10.1145/3520312.3534862} explore the capabilities of LLMs in software engineering tasks, providing a systematic evaluation of models like Codex. Their work emphasizes the strengths of LLMs in automating code generation and refactoring tasks, while also pointing out the limitations of these models in understanding deeper program semantics.

Our work differs by focusing on the code refactoring capability of LLMs like StarCoder2, comparing their performance with that of developers, specifically evaluating the reduction of code smells and improvements in code quality metrics, which has been less explored in prior research.

\subsection{Code Refactoring Using LLMs}

LLMs have also been adopted for code refactoring tasks, where the focus is on improving code quality by enhancing metrics like cyclomatic complexity and code modularity. Shirafuji et al.~\cite{10479398} demonstrate that LLM-generated refactorings can lead to significant improvements in code metrics. By generating 10 candidate solutions (pass@10) with GPT-3.5, they find 95.68\% of programs could be successfully refactored, leading to a 17.35\% reduction in cyclomatic complexity and a 25.84\% decrease in the average number of lines of code.

Our work builds on these studies by not only comparing LLM-generated refactorings with those performed by developers but also by analyzing the effectiveness of different refactoring types. We evaluate how these refactorings reduce code smells and improve key software metrics.

\subsection{Prompt Engineering and Fine-Tuning for Improving LLMs}

Prompt engineering has emerged as a technique for enhancing the performance of LLMs in various tasks, including code refactoring. Studies have shown that by carefully designing input prompts, the accuracy and quality of LLM-generated refactorings can be significantly improved. For instance, a notable study by Brown et al.~\cite{NEURIPS2020_1457c0d6} demonstrate that few-shot prompting techniques can guide LLMs to generate higher-quality code, improving overall refactoring outcomes.

Large Language Models (LLMs) have also shown abilities in code generation through Chain-of-Thought (CoT) prompting, which asks LLMs to first generate intermediate reasoning steps before outputting code. However, CoT prompting has limitations in practical applications. For example, GPT-3.5-turbo with CoT prompting achieves only 53.29\% Pass@1 in the HumanEval benchmark. Li et al.~\cite{li2023structured} propose a technique called Structured CoTs (SCoTs) to overcome this limitation. SCoT prompting introduces programming structures (i.e., sequential, branch, and loop) explicitly into the LLM’s reasoning process, unlocking structured programming thinking in the model. Evaluations on benchmarks such as HumanEval, MBPP, and MBCPP show that SCoT prompting outperforms CoT prompting by up to 13.79\% in Pass@1, and human developers prefer the programs generated by SCoT prompting over those from CoT prompting.

Fine-tuning has also been explored as a method to improve the refactoring capabilities of pre-trained models. Recent research~\cite{10.1145/3597926.3598036} has shown that task-specific fine-tuning, combined with prompt design, can further enhance the ability of LLMs to perform complex refactoring tasks. This approach leverages domain-specific data to refine the model's understanding of code structure and quality.

Our study builds on these findings by investigating how modifications to input prompts, including zero-shot, one-shot, and chain-of-thought prompting, can improve the refactoring capabilities of StarCoder2. We aim to optimize prompt designs to maximize the model's effectiveness in reducing code smells and improving code metrics.
\section{Conclusion} 
\label{sec:conclusion}
In this study we investigate the refactoring capabilities of StarCoder2, comparing its performance to human developers. Our findings indicate that StarCoder2 outperforms developers in reducing implementation code smells, achieving a reduction rate of 43.36\%, compared to developers' code smell reduction rate of 24.27\%. Moreover, StarCoder2 shows better performance when performing systematic refactorings, while developers better handle more complex, context-dependent design code smells. We observe that prompt engineering techniques, particularly one-shot prompting, significantly improve refactoring quality, emphasizing the role of prompt design in maximizing the performance of StarCoder2 in refactoring.  Our results highlight the complementary strengths of LLMs, particularly StarCoder2, and human expertise, suggesting best practices for utilizing LLMs in refactoring tasks. Future work will explore fine-tuning strategies and assess the effectiveness of LLMs across more diverse projects and programming languages.



\bibliographystyle{ACM-Reference-Format}
\bibliography{sample-manuscript}


\begin{thebibliography}{54}


\ifx \showCODEN    \undefined \def \showCODEN     #1{\unskip}     \fi
\ifx \showDOI      \undefined \def \showDOI       #1{#1}\fi
\ifx \showISBNx    \undefined \def \showISBNx     #1{\unskip}     \fi
\ifx \showISBNxiii \undefined \def \showISBNxiii  #1{\unskip}     \fi
\ifx \showISSN     \undefined \def \showISSN      #1{\unskip}     \fi
\ifx \showLCCN     \undefined \def \showLCCN      #1{\unskip}     \fi
\ifx \shownote     \undefined \def \shownote      #1{#1}          \fi
\ifx \showarticletitle \undefined \def \showarticletitle #1{#1}   \fi
\ifx \showURL      \undefined \def \showURL       {\relax}        \fi
\providecommand\bibfield[2]{#2}
\providecommand\bibinfo[2]{#2}
\providecommand\natexlab[1]{#1}
\providecommand\showeprint[2][]{arXiv:#2}

\bibitem[Alshahwan et~al\mbox{.}(2024)]%
        {10.1145/3643661.3643953}
\bibfield{author}{\bibinfo{person}{Nadia Alshahwan}, \bibinfo{person}{Mark Harman}, \bibinfo{person}{Inna Harper}, \bibinfo{person}{Alexandru Marginean}, \bibinfo{person}{Shubho Sengupta}, {and} \bibinfo{person}{Eddy Wang}.} \bibinfo{year}{2024}\natexlab{}.
\newblock \showarticletitle{Assured Offline LLM-Based Software Engineering}. In \bibinfo{booktitle}{\emph{Proceedings of the ACM/IEEE 2nd International Workshop on Interpretability, Robustness, and Benchmarking in Neural Software Engineering}} (Lisbon, Portugal) \emph{(\bibinfo{series}{InteNSE '24})}. \bibinfo{publisher}{Association for Computing Machinery}, \bibinfo{address}{New York, NY, USA}, \bibinfo{pages}{7–12}.
\newblock
\showISBNx{9798400705649}
\urldef\tempurl%
\url{https://doi.org/10.1145/3643661.3643953}
\showDOI{\tempurl}


\bibitem[Balog et~al\mbox{.}(2016)]%
        {DBLP:journals/corr/BalogGBNT16}
\bibfield{author}{\bibinfo{person}{Matej Balog}, \bibinfo{person}{Alexander~L. Gaunt}, \bibinfo{person}{Marc Brockschmidt}, \bibinfo{person}{Sebastian Nowozin}, {and} \bibinfo{person}{Daniel Tarlow}.} \bibinfo{year}{2016}\natexlab{}.
\newblock \showarticletitle{DeepCoder: Learning to Write Programs}.
\newblock \bibinfo{journal}{\emph{CoRR}}  \bibinfo{volume}{abs/1611.01989} (\bibinfo{year}{2016}).
\newblock
\showeprint[arXiv]{1611.01989}
\urldef\tempurl%
\url{http://arxiv.org/abs/1611.01989}
\showURL{%
\tempurl}


\bibitem[{Berkeley Bootcamp}(2020)]%
        {berkeley2020indemand}
\bibfield{author}{\bibinfo{person}{{Berkeley Bootcamp}}.} \bibinfo{year}{2020}\natexlab{}.
\newblock \bibinfo{title}{11 Most In-Demand Programming Languages in 2021}.
\newblock
\newblock
\newblock
\shownote{[Online]. Available: \url{https://bootcamp.berkeley.edu/blog/most-indemand-programming-languages/}}.


\bibitem[Brown et~al\mbox{.}(2020)]%
        {NEURIPS2020_1457c0d6}
\bibfield{author}{\bibinfo{person}{Tom Brown}, \bibinfo{person}{Benjamin Mann}, \bibinfo{person}{Nick Ryder}, \bibinfo{person}{Melanie Subbiah}, \bibinfo{person}{Jared~D Kaplan}, \bibinfo{person}{Prafulla Dhariwal}, \bibinfo{person}{Arvind Neelakantan}, \bibinfo{person}{Pranav Shyam}, \bibinfo{person}{Girish Sastry}, \bibinfo{person}{Amanda Askell}, \bibinfo{person}{Sandhini Agarwal}, \bibinfo{person}{Ariel Herbert-Voss}, \bibinfo{person}{Gretchen Krueger}, \bibinfo{person}{Tom Henighan}, \bibinfo{person}{Rewon Child}, \bibinfo{person}{Aditya Ramesh}, \bibinfo{person}{Daniel Ziegler}, \bibinfo{person}{Jeffrey Wu}, \bibinfo{person}{Clemens Winter}, \bibinfo{person}{Chris Hesse}, \bibinfo{person}{Mark Chen}, \bibinfo{person}{Eric Sigler}, \bibinfo{person}{Mateusz Litwin}, \bibinfo{person}{Scott Gray}, \bibinfo{person}{Benjamin Chess}, \bibinfo{person}{Jack Clark}, \bibinfo{person}{Christopher Berner}, \bibinfo{person}{Sam McCandlish}, \bibinfo{person}{Alec Radford}, \bibinfo{person}{Ilya Sutskever}, {and}
  \bibinfo{person}{Dario Amodei}.} \bibinfo{year}{2020}\natexlab{}.
\newblock \showarticletitle{Language Models are Few-Shot Learners}. In \bibinfo{booktitle}{\emph{Advances in Neural Information Processing Systems}}, \bibfield{editor}{\bibinfo{person}{H.~Larochelle}, \bibinfo{person}{M.~Ranzato}, \bibinfo{person}{R.~Hadsell}, \bibinfo{person}{M.F. Balcan}, {and} \bibinfo{person}{H.~Lin}} (Eds.), Vol.~\bibinfo{volume}{33}. \bibinfo{publisher}{Curran Associates, Inc.}, \bibinfo{pages}{1877--1901}.
\newblock
\urldef\tempurl%
\url{https://proceedings.neurips.cc/paper_files/paper/2020/file/1457c0d6bfcb4967418bfb8ac142f64a-Paper.pdf}
\showURL{%
\tempurl}


\bibitem[Brownlee(2020)]%
        {Brownlee_2020}
\bibfield{author}{\bibinfo{person}{Jason Brownlee}.} \bibinfo{year}{2020}\natexlab{}.
\newblock \bibinfo{title}{Data leakage in machine learning}.
\newblock
\newblock
\urldef\tempurl%
\url{https://machinelearningmastery.com/data-leakage-machine-learning/}
\showURL{%
\tempurl}


\bibitem[Cedrim et~al\mbox{.}(2017)]%
        {10.1145/3106237.3106259}
\bibfield{author}{\bibinfo{person}{Diego Cedrim}, \bibinfo{person}{Alessandro Garcia}, \bibinfo{person}{Melina Mongiovi}, \bibinfo{person}{Rohit Gheyi}, \bibinfo{person}{Leonardo Sousa}, \bibinfo{person}{Rafael de Mello}, \bibinfo{person}{Baldoino Fonseca}, \bibinfo{person}{M\'{a}rcio Ribeiro}, {and} \bibinfo{person}{Alexander Ch\'{a}vez}.} \bibinfo{year}{2017}\natexlab{}.
\newblock \showarticletitle{Understanding the impact of refactoring on smells: a longitudinal study of 23 software projects}. In \bibinfo{booktitle}{\emph{Proceedings of the 2017 11th Joint Meeting on Foundations of Software Engineering}} (Paderborn, Germany) \emph{(\bibinfo{series}{ESEC/FSE 2017})}. \bibinfo{publisher}{Association for Computing Machinery}, \bibinfo{address}{New York, NY, USA}, \bibinfo{pages}{465–475}.
\newblock
\showISBNx{9781450351058}
\urldef\tempurl%
\url{https://doi.org/10.1145/3106237.3106259}
\showDOI{\tempurl}


\bibitem[Chang et~al\mbox{.}(2024)]%
        {chang2024survey}
\bibfield{author}{\bibinfo{person}{Yupeng Chang}, \bibinfo{person}{Xu Wang}, \bibinfo{person}{Jindong Wang}, \bibinfo{person}{Yuan Wu}, \bibinfo{person}{Linyi Yang}, \bibinfo{person}{Kaijie Zhu}, \bibinfo{person}{Hao Chen}, \bibinfo{person}{Xiaoyuan Yi}, \bibinfo{person}{Cunxiang Wang}, \bibinfo{person}{Yidong Wang}, {et~al\mbox{.}}} \bibinfo{year}{2024}\natexlab{}.
\newblock \showarticletitle{A survey on evaluation of large language models}.
\newblock \bibinfo{journal}{\emph{ACM Transactions on Intelligent Systems and Technology}} \bibinfo{volume}{15}, \bibinfo{number}{3} (\bibinfo{year}{2024}), \bibinfo{pages}{1--45}.
\newblock


\bibitem[Choi et~al\mbox{.}(2024)]%
        {Choi2024aa}
\bibfield{author}{\bibinfo{person}{Jinsu Choi}, \bibinfo{person}{Gabin An}, {and} \bibinfo{person}{Shin Yoo}.} \bibinfo{year}{2024}\natexlab{}.
\newblock \showarticletitle{Iterative Refactoring of Real-World Open-Source Programs with Large Language Models}.
\newblock In \bibinfo{booktitle}{\emph{Proceedings of the 16th International Symposium on Search-Based Software Engineering}}. \bibinfo{series}{Lecture Notes in Computer Science}, Vol.~\bibinfo{volume}{14767}. \bibinfo{publisher}{Springer Nature}, \bibinfo{pages}{49--55}.
\newblock


\bibitem[Chowdhery et~al\mbox{.}(2022)]%
        {chowdhery2022palmscalinglanguagemodeling}
\bibfield{author}{\bibinfo{person}{Aakanksha Chowdhery}, \bibinfo{person}{Sharan Narang}, \bibinfo{person}{Jacob Devlin}, \bibinfo{person}{Maarten Bosma}, \bibinfo{person}{Gaurav Mishra}, \bibinfo{person}{Adam Roberts}, \bibinfo{person}{Paul Barham}, \bibinfo{person}{Hyung~Won Chung}, \bibinfo{person}{Charles Sutton}, \bibinfo{person}{Sebastian Gehrmann}, \bibinfo{person}{Parker Schuh}, \bibinfo{person}{Kensen Shi}, \bibinfo{person}{Sasha Tsvyashchenko}, \bibinfo{person}{Joshua Maynez}, \bibinfo{person}{Abhishek Rao}, \bibinfo{person}{Parker Barnes}, \bibinfo{person}{Yi Tay}, \bibinfo{person}{Noam Shazeer}, \bibinfo{person}{Vinodkumar Prabhakaran}, \bibinfo{person}{Emily Reif}, \bibinfo{person}{Nan Du}, \bibinfo{person}{Ben Hutchinson}, \bibinfo{person}{Reiner Pope}, \bibinfo{person}{James Bradbury}, \bibinfo{person}{Jacob Austin}, \bibinfo{person}{Michael Isard}, \bibinfo{person}{Guy Gur-Ari}, \bibinfo{person}{Pengcheng Yin}, \bibinfo{person}{Toju Duke}, \bibinfo{person}{Anselm Levskaya},
  \bibinfo{person}{Sanjay Ghemawat}, \bibinfo{person}{Sunipa Dev}, \bibinfo{person}{Henryk Michalewski}, \bibinfo{person}{Xavier Garcia}, \bibinfo{person}{Vedant Misra}, \bibinfo{person}{Kevin Robinson}, \bibinfo{person}{Liam Fedus}, \bibinfo{person}{Denny Zhou}, \bibinfo{person}{Daphne Ippolito}, \bibinfo{person}{David Luan}, \bibinfo{person}{Hyeontaek Lim}, \bibinfo{person}{Barret Zoph}, \bibinfo{person}{Alexander Spiridonov}, \bibinfo{person}{Ryan Sepassi}, \bibinfo{person}{David Dohan}, \bibinfo{person}{Shivani Agrawal}, \bibinfo{person}{Mark Omernick}, \bibinfo{person}{Andrew~M. Dai}, \bibinfo{person}{Thanumalayan~Sankaranarayana Pillai}, \bibinfo{person}{Marie Pellat}, \bibinfo{person}{Aitor Lewkowycz}, \bibinfo{person}{Erica Moreira}, \bibinfo{person}{Rewon Child}, \bibinfo{person}{Oleksandr Polozov}, \bibinfo{person}{Katherine Lee}, \bibinfo{person}{Zongwei Zhou}, \bibinfo{person}{Xuezhi Wang}, \bibinfo{person}{Brennan Saeta}, \bibinfo{person}{Mark Diaz}, \bibinfo{person}{Orhan Firat},
  \bibinfo{person}{Michele Catasta}, \bibinfo{person}{Jason Wei}, \bibinfo{person}{Kathy Meier-Hellstern}, \bibinfo{person}{Douglas Eck}, \bibinfo{person}{Jeff Dean}, \bibinfo{person}{Slav Petrov}, {and} \bibinfo{person}{Noah Fiedel}.} \bibinfo{year}{2022}\natexlab{}.
\newblock \bibinfo{title}{PaLM: Scaling Language Modeling with Pathways}.
\newblock
\newblock
\showeprint[arxiv]{2204.02311}~[cs.CL]
\urldef\tempurl%
\url{https://arxiv.org/abs/2204.02311}
\showURL{%
\tempurl}


\bibitem[Claes and M{\"a}ntyl{\"a}(2020)]%
        {claes2020mad}
\bibfield{author}{\bibinfo{person}{M. Claes} {and} \bibinfo{person}{M.~V. M{\"a}ntyl{\"a}}.} \bibinfo{year}{2020}\natexlab{}.
\newblock \showarticletitle{20-mad: 20 years of issues and commits of mozilla and apache development}. In \bibinfo{booktitle}{\emph{Proceedings of the 17th International Conference on Mining Software Repositories}}. \bibinfo{pages}{503--507}.
\newblock


\bibitem[{Enio G. Jelihovschi} et~al\mbox{.}(2014)]%
        {scott-knott}
\bibfield{author}{\bibinfo{person}{{Enio G. Jelihovschi}}, \bibinfo{person}{{Jose Claudio Faria}}, {and} \bibinfo{person}{{Ivan Bezerra Allaman}}.} \bibinfo{year}{2014}\natexlab{}.
\newblock \showarticletitle{ScottKnott: A Package for Performing the Scott-Knott Clustering Algorithm in R.}
\newblock \bibinfo{journal}{\emph{Trends in Applied and Computational Mathematics}} \bibinfo{volume}{15}, \bibinfo{number}{1} (\bibinfo{year}{2014}), \bibinfo{pages}{3--17}.
\newblock
\urldef\tempurl%
\url{https://tema.sbmac.org.br/tema/article/view/646/643}
\showURL{%
\tempurl}


\bibitem[Fan et~al\mbox{.}(2023b)]%
        {fan2023large}
\bibfield{author}{\bibinfo{person}{Angela Fan}, \bibinfo{person}{Beliz Gokkaya}, \bibinfo{person}{Mark Harman}, \bibinfo{person}{Mitya Lyubarskiy}, \bibinfo{person}{Shubho Sengupta}, \bibinfo{person}{Shin Yoo}, {and} \bibinfo{person}{Jie~M Zhang}.} \bibinfo{year}{2023}\natexlab{b}.
\newblock \showarticletitle{Large language models for software engineering: Survey and open problems}. In \bibinfo{booktitle}{\emph{2023 IEEE/ACM International Conference on Software Engineering: Future of Software Engineering (ICSE-FoSE)}}. IEEE, \bibinfo{pages}{31--53}.
\newblock


\bibitem[Fan et~al\mbox{.}(2023a)]%
        {10.1109/ICSE48619.2023.00128}
\bibfield{author}{\bibinfo{person}{Zhiyu Fan}, \bibinfo{person}{Xiang Gao}, \bibinfo{person}{Martin Mirchev}, \bibinfo{person}{Abhik Roychoudhury}, {and} \bibinfo{person}{Shin~Hwei Tan}.} \bibinfo{year}{2023}\natexlab{a}.
\newblock \showarticletitle{Automated Repair of Programs from Large Language Models}. In \bibinfo{booktitle}{\emph{Proceedings of the 45th International Conference on Software Engineering}} (Melbourne, Victoria, Australia) \emph{(\bibinfo{series}{ICSE '23})}. \bibinfo{publisher}{IEEE Press}, \bibinfo{pages}{1469–1481}.
\newblock
\showISBNx{9781665457019}
\urldef\tempurl%
\url{https://doi.org/10.1109/ICSE48619.2023.00128}
\showDOI{\tempurl}


\bibitem[Fowler(2018)]%
        {fowler2018refactoring}
\bibfield{author}{\bibinfo{person}{Martin Fowler}.} \bibinfo{year}{2018}\natexlab{}.
\newblock \bibinfo{booktitle}{\emph{Refactoring: improving the design of existing code}}.
\newblock \bibinfo{publisher}{Addison-Wesley Professional}.
\newblock


\bibitem[Fowler et~al\mbox{.}(1999)]%
        {fowler1999refactoring}
\bibfield{author}{\bibinfo{person}{M. Fowler}, \bibinfo{person}{K. Beck}, \bibinfo{person}{J. Brant}, \bibinfo{person}{W. Opdyke}, {and} \bibinfo{person}{D. Roberts}.} \bibinfo{year}{1999}\natexlab{}.
\newblock \bibinfo{booktitle}{\emph{Refactoring: Improving the Design of Existing Code}}.
\newblock \bibinfo{publisher}{Addison-Wesley Professional}, \bibinfo{address}{Berkeley, CA, USA}.
\newblock


\bibitem[Fraser and Arcuri(2011)]%
        {inproceedings}
\bibfield{author}{\bibinfo{person}{Gordon Fraser} {and} \bibinfo{person}{Andrea Arcuri}.} \bibinfo{year}{2011}\natexlab{}.
\newblock \showarticletitle{EvoSuite: Automatic test suite generation for object-oriented software}.
\newblock \bibinfo{journal}{\emph{SIGSOFT/FSE 2011 - Proceedings of the 19th ACM SIGSOFT Symposium on Foundations of Software Engineering}}, \bibinfo{pages}{416--419}.
\newblock
\urldef\tempurl%
\url{https://doi.org/10.1145/2025113.2025179}
\showDOI{\tempurl}


\bibitem[Guru(2024)]%
        {Refactoring.Guru}
\bibfield{author}{\bibinfo{person}{Refactoring Guru}.} \bibinfo{year}{2024}\natexlab{}.
\newblock
\newblock
\urldef\tempurl%
\url{https://refactoring.guru/refactoring/techniques}
\showURL{%
\tempurl}


\bibitem[Hazra(2017)]%
        {hazra2017using}
\bibfield{author}{\bibinfo{person}{Avijit Hazra}.} \bibinfo{year}{2017}\natexlab{}.
\newblock \showarticletitle{Using the confidence interval confidently}.
\newblock \bibinfo{journal}{\emph{Journal of thoracic disease}} \bibinfo{volume}{9}, \bibinfo{number}{10} (\bibinfo{year}{2017}), \bibinfo{pages}{4125}.
\newblock


\bibitem[Jansen(2024)]%
        {TIOBE_2022}
\bibfield{author}{\bibinfo{person}{Paul Jansen}.} \bibinfo{year}{2024}\natexlab{}.
\newblock
\newblock
\urldef\tempurl%
\url{https://www.tiobe.com/tiobe-index/}
\showURL{%
\tempurl}


\bibitem[Kojima et~al\mbox{.}(2022)]%
        {NEURIPS2022_8bb0d291}
\bibfield{author}{\bibinfo{person}{Takeshi Kojima}, \bibinfo{person}{Shixiang~(Shane) Gu}, \bibinfo{person}{Machel Reid}, \bibinfo{person}{Yutaka Matsuo}, {and} \bibinfo{person}{Yusuke Iwasawa}.} \bibinfo{year}{2022}\natexlab{}.
\newblock \showarticletitle{Large Language Models are Zero-Shot Reasoners}. In \bibinfo{booktitle}{\emph{Advances in Neural Information Processing Systems}}, \bibfield{editor}{\bibinfo{person}{S.~Koyejo}, \bibinfo{person}{S.~Mohamed}, \bibinfo{person}{A.~Agarwal}, \bibinfo{person}{D.~Belgrave}, \bibinfo{person}{K.~Cho}, {and} \bibinfo{person}{A.~Oh}} (Eds.), Vol.~\bibinfo{volume}{35}. \bibinfo{publisher}{Curran Associates, Inc.}, \bibinfo{pages}{22199--22213}.
\newblock
\urldef\tempurl%
\url{https://proceedings.neurips.cc/paper_files/paper/2022/file/8bb0d291acd4acf06ef112099c16f326-Paper-Conference.pdf}
\showURL{%
\tempurl}


\bibitem[Li et~al\mbox{.}(2023b)]%
        {li2023structured}
\bibfield{author}{\bibinfo{person}{Jia Li}, \bibinfo{person}{Ge Li}, \bibinfo{person}{Yongmin Li}, {and} \bibinfo{person}{Zhi Jin}.} \bibinfo{year}{2023}\natexlab{b}.
\newblock \showarticletitle{Structured chain-of-thought prompting for code generation}.
\newblock \bibinfo{journal}{\emph{ACM Transactions on Software Engineering and Methodology}} (\bibinfo{year}{2023}).
\newblock


\bibitem[Li et~al\mbox{.}(2023a)]%
        {li2023starcoder}
\bibfield{author}{\bibinfo{person}{Raymond Li}, \bibinfo{person}{Loubna~Ben Allal}, \bibinfo{person}{Yangtian Zi}, \bibinfo{person}{Niklas Muennighoff}, \bibinfo{person}{Denis Kocetkov}, \bibinfo{person}{Chenghao Mou}, \bibinfo{person}{Marc Marone}, \bibinfo{person}{Christopher Akiki}, \bibinfo{person}{Jia Li}, \bibinfo{person}{Jenny Chim}, \bibinfo{person}{Qian Liu}, \bibinfo{person}{Evgenii Zheltonozhskii}, \bibinfo{person}{Terry~Yue Zhuo}, \bibinfo{person}{Thomas Wang}, \bibinfo{person}{Olivier Dehaene}, \bibinfo{person}{Mishig Davaadorj}, \bibinfo{person}{Joel Lamy-Poirier}, \bibinfo{person}{João Monteiro}, \bibinfo{person}{Oleh Shliazhko}, \bibinfo{person}{Nicolas Gontier}, \bibinfo{person}{Nicholas Meade}, \bibinfo{person}{Armel Zebaze}, \bibinfo{person}{Ming-Ho Yee}, \bibinfo{person}{Logesh~Kumar Umapathi}, \bibinfo{person}{Jian Zhu}, \bibinfo{person}{Benjamin Lipkin}, \bibinfo{person}{Muhtasham Oblokulov}, \bibinfo{person}{Zhiruo Wang}, \bibinfo{person}{Rudra Murthy}, \bibinfo{person}{Jason
  Stillerman}, \bibinfo{person}{Siva~Sankalp Patel}, \bibinfo{person}{Dmitry Abulkhanov}, \bibinfo{person}{Marco Zocca}, \bibinfo{person}{Manan Dey}, \bibinfo{person}{Zhihan Zhang}, \bibinfo{person}{Nour Fahmy}, \bibinfo{person}{Urvashi Bhattacharyya}, \bibinfo{person}{Wenhao Yu}, \bibinfo{person}{Swayam Singh}, \bibinfo{person}{Sasha Luccioni}, \bibinfo{person}{Paulo Villegas}, \bibinfo{person}{Maxim Kunakov}, \bibinfo{person}{Fedor Zhdanov}, \bibinfo{person}{Manuel Romero}, \bibinfo{person}{Tony Lee}, \bibinfo{person}{Nadav Timor}, \bibinfo{person}{Jennifer Ding}, \bibinfo{person}{Claire Schlesinger}, \bibinfo{person}{Hailey Schoelkopf}, \bibinfo{person}{Jan Ebert}, \bibinfo{person}{Tri Dao}, \bibinfo{person}{Mayank Mishra}, \bibinfo{person}{Alex Gu}, \bibinfo{person}{Jennifer Robinson}, \bibinfo{person}{Carolyn~Jane Anderson}, \bibinfo{person}{Brendan Dolan-Gavitt}, \bibinfo{person}{Danish Contractor}, \bibinfo{person}{Siva Reddy}, \bibinfo{person}{Daniel Fried}, \bibinfo{person}{Dzmitry Bahdanau},
  \bibinfo{person}{Yacine Jernite}, \bibinfo{person}{Carlos~Muñoz Ferrandis}, \bibinfo{person}{Sean Hughes}, \bibinfo{person}{Thomas Wolf}, \bibinfo{person}{Arjun Guha}, \bibinfo{person}{Leandro von Werra}, {and} \bibinfo{person}{Harm de Vries}.} \bibinfo{year}{2023}\natexlab{a}.
\newblock \bibinfo{title}{StarCoder: may the source be with you!}
\newblock
\newblock
\showeprint[arxiv]{2305.06161}~[cs.CL]


\bibitem[Li et~al\mbox{.}(2022)]%
        {li2022competition}
\bibfield{author}{\bibinfo{person}{Yujia Li}, \bibinfo{person}{David Choi}, \bibinfo{person}{Junyoung Chung}, \bibinfo{person}{Nate Kushman}, \bibinfo{person}{Julian Schrittwieser}, \bibinfo{person}{R{\'e}mi Leblond}, \bibinfo{person}{Tom Eccles}, \bibinfo{person}{James Keeling}, \bibinfo{person}{Felix Gimeno}, \bibinfo{person}{Agustin Dal~Lago}, {et~al\mbox{.}}} \bibinfo{year}{2022}\natexlab{}.
\newblock \showarticletitle{Competition-level code generation with alphacode}.
\newblock \bibinfo{journal}{\emph{Science}} \bibinfo{volume}{378}, \bibinfo{number}{6624} (\bibinfo{year}{2022}), \bibinfo{pages}{1092--1097}.
\newblock


\bibitem[Lozhkov et~al\mbox{.}(2024)]%
        {lozhkov2024starcoder}
\bibfield{author}{\bibinfo{person}{Anton Lozhkov}, \bibinfo{person}{Raymond Li}, \bibinfo{person}{Loubna~Ben Allal}, \bibinfo{person}{Federico Cassano}, \bibinfo{person}{Joel Lamy-Poirier}, \bibinfo{person}{Nouamane Tazi}, \bibinfo{person}{Ao Tang}, \bibinfo{person}{Dmytro Pykhtar}, \bibinfo{person}{Jiawei Liu}, \bibinfo{person}{Yuxiang Wei}, {et~al\mbox{.}}} \bibinfo{year}{2024}\natexlab{}.
\newblock \showarticletitle{Starcoder 2 and the stack v2: The next generation}.
\newblock \bibinfo{journal}{\emph{arXiv preprint arXiv:2402.19173}} (\bibinfo{year}{2024}).
\newblock


\bibitem[McKnight and Najab(2010)]%
        {doi:https://doi.org/10.1002/9780470479216.corpsy0524}
\bibfield{author}{\bibinfo{person}{Patrick~E. McKnight} {and} \bibinfo{person}{Julius Najab}.} \bibinfo{year}{2010}\natexlab{}.
\newblock \bibinfo{booktitle}{\emph{Mann-Whitney U Test}}.
\newblock \bibinfo{publisher}{John Wiley I\& Sons, Ltd}, \bibinfo{pages}{1--1}.
\newblock
\showISBNx{9780470479216}
\urldef\tempurl%
\url{https://doi.org/10.1002/9780470479216.corpsy0524}
\showDOI{\tempurl}
\showeprint{https://onlinelibrary.wiley.com/doi/pdf/10.1002/9780470479216.corpsy0524}


\bibitem[Meissel and Yao(2024)]%
        {meissel2024cliffs}
\bibfield{author}{\bibinfo{person}{Katherine Meissel} {and} \bibinfo{person}{Ernest~S. Yao}.} \bibinfo{year}{2024}\natexlab{}.
\newblock \showarticletitle{Using Cliff’s Delta as a Non-Parametric Effect Size Measure: An Accessible Web App and R Tutorial}.
\newblock \bibinfo{journal}{\emph{Practical Assessment, Research, and Evaluation}} \bibinfo{volume}{29}, \bibinfo{number}{1} (\bibinfo{year}{2024}), \bibinfo{pages}{2}.
\newblock
\urldef\tempurl%
\url{https://doi.org/10.7275/pare.1977}
\showDOI{\tempurl}


\bibitem[Mens and Tourwe(2004)]%
        {1265817}
\bibfield{author}{\bibinfo{person}{T. Mens} {and} \bibinfo{person}{T. Tourwe}.} \bibinfo{year}{2004}\natexlab{}.
\newblock \showarticletitle{A survey of software refactoring}.
\newblock \bibinfo{journal}{\emph{IEEE Transactions on Software Engineering}} \bibinfo{volume}{30}, \bibinfo{number}{2} (\bibinfo{year}{2004}), \bibinfo{pages}{126--139}.
\newblock
\urldef\tempurl%
\url{https://doi.org/10.1109/TSE.2004.1265817}
\showDOI{\tempurl}


\bibitem[Miller et~al\mbox{.}(2010)]%
        {miller2010apache}
\bibfield{author}{\bibinfo{person}{Frederic~P Miller}, \bibinfo{person}{Agnes~F Vandome}, {and} \bibinfo{person}{John McBrewster}.} \bibinfo{year}{2010}\natexlab{}.
\newblock \bibinfo{booktitle}{\emph{Apache Maven}}.
\newblock \bibinfo{publisher}{Alpha Press}.
\newblock


\bibitem[mockito(2024)]%
        {mockitoMockitoFramework}
\bibfield{author}{\bibinfo{person}{mockito}.} \bibinfo{year}{2024}\natexlab{}.
\newblock
\newblock
\urldef\tempurl%
\url{https://site.mockito.org/}
\showURL{%
\tempurl}


\bibitem[Noei et~al\mbox{.}(2023)]%
        {noei2023empirical}
\bibfield{author}{\bibinfo{person}{Shayan Noei}, \bibinfo{person}{Heng Li}, \bibinfo{person}{Stefanos Georgiou}, {and} \bibinfo{person}{Ying Zou}.} \bibinfo{year}{2023}\natexlab{}.
\newblock \showarticletitle{An Empirical Study of Refactoring Rhythms and Tactics in the Software Development Process}.
\newblock \bibinfo{journal}{\emph{IEEE Transactions on Software Engineering}} \bibinfo{number}{01} (\bibinfo{year}{2023}), \bibinfo{pages}{1--17}.
\newblock


\bibitem[Nu{\~n}ez-Varela et~al\mbox{.}(2017)]%
        {nunez2017source}
\bibfield{author}{\bibinfo{person}{Alberto~S Nu{\~n}ez-Varela}, \bibinfo{person}{H{\'e}ctor~G P{\'e}rez-Gonzalez}, \bibinfo{person}{Francisco~E Mart{\'\i}nez-Perez}, {and} \bibinfo{person}{Carlos Soubervielle-Montalvo}.} \bibinfo{year}{2017}\natexlab{}.
\newblock \showarticletitle{Source code metrics: A systematic mapping study}.
\newblock \bibinfo{journal}{\emph{Journal of Systems and Software}}  \bibinfo{volume}{128} (\bibinfo{year}{2017}), \bibinfo{pages}{164--197}.
\newblock


\bibitem[Opdyke and Johnson(1992)]%
        {article}
\bibfield{author}{\bibinfo{person}{William Opdyke} {and} \bibinfo{person}{Ralph Johnson}.} \bibinfo{year}{1992}\natexlab{}.
\newblock \showarticletitle{Refactoring Object-Oriented Frameworks}.
\newblock  (\bibinfo{date}{07} \bibinfo{year}{1992}).
\newblock


\bibitem[OpenAI et~al\mbox{.}(2024)]%
        {openai2024gpt4}
\bibfield{author}{\bibinfo{person}{OpenAI}, \bibinfo{person}{Josh Achiam}, \bibinfo{person}{Steven Adler}, \bibinfo{person}{Sandhini Agarwal}, \bibinfo{person}{Lama Ahmad}, \bibinfo{person}{Ilge Akkaya}, \bibinfo{person}{Florencia~Leoni Aleman}, \bibinfo{person}{Diogo Almeida}, \bibinfo{person}{Janko Altenschmidt}, \bibinfo{person}{Sam Altman}, \bibinfo{person}{Shyamal Anadkat}, \bibinfo{person}{Red Avila}, \bibinfo{person}{Igor Babuschkin}, \bibinfo{person}{Suchir Balaji}, \bibinfo{person}{Valerie Balcom}, \bibinfo{person}{Paul Baltescu}, \bibinfo{person}{Haiming Bao}, \bibinfo{person}{Mohammad Bavarian}, \bibinfo{person}{Jeff Belgum}, \bibinfo{person}{Irwan Bello}, \bibinfo{person}{Jake Berdine}, \bibinfo{person}{Gabriel Bernadett-Shapiro}, \bibinfo{person}{Christopher Berner}, \bibinfo{person}{Lenny Bogdonoff}, \bibinfo{person}{Oleg Boiko}, \bibinfo{person}{Madelaine Boyd}, \bibinfo{person}{Anna-Luisa Brakman}, \bibinfo{person}{Greg Brockman}, \bibinfo{person}{Tim Brooks}, \bibinfo{person}{Miles Brundage},
  \bibinfo{person}{Kevin Button}, \bibinfo{person}{Trevor Cai}, \bibinfo{person}{Rosie Campbell}, \bibinfo{person}{Andrew Cann}, \bibinfo{person}{Brittany Carey}, \bibinfo{person}{Chelsea Carlson}, \bibinfo{person}{Rory Carmichael}, \bibinfo{person}{Brooke Chan}, \bibinfo{person}{Che Chang}, \bibinfo{person}{Fotis Chantzis}, \bibinfo{person}{Derek Chen}, \bibinfo{person}{Sully Chen}, \bibinfo{person}{Ruby Chen}, \bibinfo{person}{Jason Chen}, \bibinfo{person}{Mark Chen}, \bibinfo{person}{Ben Chess}, \bibinfo{person}{Chester Cho}, \bibinfo{person}{Casey Chu}, \bibinfo{person}{Hyung~Won Chung}, \bibinfo{person}{Dave Cummings}, \bibinfo{person}{Jeremiah Currier}, \bibinfo{person}{Yunxing Dai}, \bibinfo{person}{Cory Decareaux}, \bibinfo{person}{Thomas Degry}, \bibinfo{person}{Noah Deutsch}, \bibinfo{person}{Damien Deville}, \bibinfo{person}{Arka Dhar}, \bibinfo{person}{David Dohan}, \bibinfo{person}{Steve Dowling}, \bibinfo{person}{Sheila Dunning}, \bibinfo{person}{Adrien Ecoffet}, \bibinfo{person}{Atty Eleti},
  \bibinfo{person}{Tyna Eloundou}, \bibinfo{person}{David Farhi}, \bibinfo{person}{Liam Fedus}, \bibinfo{person}{Niko Felix}, \bibinfo{person}{Simón~Posada Fishman}, \bibinfo{person}{Juston Forte}, \bibinfo{person}{Isabella Fulford}, \bibinfo{person}{Leo Gao}, \bibinfo{person}{Elie Georges}, \bibinfo{person}{Christian Gibson}, \bibinfo{person}{Vik Goel}, \bibinfo{person}{Tarun Gogineni}, \bibinfo{person}{Gabriel Goh}, \bibinfo{person}{Rapha Gontijo-Lopes}, \bibinfo{person}{Jonathan Gordon}, \bibinfo{person}{Morgan Grafstein}, \bibinfo{person}{Scott Gray}, \bibinfo{person}{Ryan Greene}, \bibinfo{person}{Joshua Gross}, \bibinfo{person}{Shixiang~Shane Gu}, \bibinfo{person}{Yufei Guo}, \bibinfo{person}{Chris Hallacy}, \bibinfo{person}{Jesse Han}, \bibinfo{person}{Jeff Harris}, \bibinfo{person}{Yuchen He}, \bibinfo{person}{Mike Heaton}, \bibinfo{person}{Johannes Heidecke}, \bibinfo{person}{Chris Hesse}, \bibinfo{person}{Alan Hickey}, \bibinfo{person}{Wade Hickey}, \bibinfo{person}{Peter Hoeschele},
  \bibinfo{person}{Brandon Houghton}, \bibinfo{person}{Kenny Hsu}, \bibinfo{person}{Shengli Hu}, \bibinfo{person}{Xin Hu}, \bibinfo{person}{Joost Huizinga}, \bibinfo{person}{Shantanu Jain}, \bibinfo{person}{Shawn Jain}, \bibinfo{person}{Joanne Jang}, \bibinfo{person}{Angela Jiang}, \bibinfo{person}{Roger Jiang}, \bibinfo{person}{Haozhun Jin}, \bibinfo{person}{Denny Jin}, \bibinfo{person}{Shino Jomoto}, \bibinfo{person}{Billie Jonn}, \bibinfo{person}{Heewoo Jun}, \bibinfo{person}{Tomer Kaftan}, \bibinfo{person}{Łukasz Kaiser}, \bibinfo{person}{Ali Kamali}, \bibinfo{person}{Ingmar Kanitscheider}, \bibinfo{person}{Nitish~Shirish Keskar}, \bibinfo{person}{Tabarak Khan}, \bibinfo{person}{Logan Kilpatrick}, \bibinfo{person}{Jong~Wook Kim}, \bibinfo{person}{Christina Kim}, \bibinfo{person}{Yongjik Kim}, \bibinfo{person}{Jan~Hendrik Kirchner}, \bibinfo{person}{Jamie Kiros}, \bibinfo{person}{Matt Knight}, \bibinfo{person}{Daniel Kokotajlo}, \bibinfo{person}{Łukasz Kondraciuk}, \bibinfo{person}{Andrew Kondrich},
  \bibinfo{person}{Aris Konstantinidis}, \bibinfo{person}{Kyle Kosic}, \bibinfo{person}{Gretchen Krueger}, \bibinfo{person}{Vishal Kuo}, \bibinfo{person}{Michael Lampe}, \bibinfo{person}{Ikai Lan}, \bibinfo{person}{Teddy Lee}, \bibinfo{person}{Jan Leike}, \bibinfo{person}{Jade Leung}, \bibinfo{person}{Daniel Levy}, \bibinfo{person}{Chak~Ming Li}, \bibinfo{person}{Rachel Lim}, \bibinfo{person}{Molly Lin}, \bibinfo{person}{Stephanie Lin}, \bibinfo{person}{Mateusz Litwin}, \bibinfo{person}{Theresa Lopez}, \bibinfo{person}{Ryan Lowe}, \bibinfo{person}{Patricia Lue}, \bibinfo{person}{Anna Makanju}, \bibinfo{person}{Kim Malfacini}, \bibinfo{person}{Sam Manning}, \bibinfo{person}{Todor Markov}, \bibinfo{person}{Yaniv Markovski}, \bibinfo{person}{Bianca Martin}, \bibinfo{person}{Katie Mayer}, \bibinfo{person}{Andrew Mayne}, \bibinfo{person}{Bob McGrew}, \bibinfo{person}{Scott~Mayer McKinney}, \bibinfo{person}{Christine McLeavey}, \bibinfo{person}{Paul McMillan}, \bibinfo{person}{Jake McNeil}, \bibinfo{person}{David
  Medina}, \bibinfo{person}{Aalok Mehta}, \bibinfo{person}{Jacob Menick}, \bibinfo{person}{Luke Metz}, \bibinfo{person}{Andrey Mishchenko}, \bibinfo{person}{Pamela Mishkin}, \bibinfo{person}{Vinnie Monaco}, \bibinfo{person}{Evan Morikawa}, \bibinfo{person}{Daniel Mossing}, \bibinfo{person}{Tong Mu}, \bibinfo{person}{Mira Murati}, \bibinfo{person}{Oleg Murk}, \bibinfo{person}{David Mély}, \bibinfo{person}{Ashvin Nair}, \bibinfo{person}{Reiichiro Nakano}, \bibinfo{person}{Rajeev Nayak}, \bibinfo{person}{Arvind Neelakantan}, \bibinfo{person}{Richard Ngo}, \bibinfo{person}{Hyeonwoo Noh}, \bibinfo{person}{Long Ouyang}, \bibinfo{person}{Cullen O'Keefe}, \bibinfo{person}{Jakub Pachocki}, \bibinfo{person}{Alex Paino}, \bibinfo{person}{Joe Palermo}, \bibinfo{person}{Ashley Pantuliano}, \bibinfo{person}{Giambattista Parascandolo}, \bibinfo{person}{Joel Parish}, \bibinfo{person}{Emy Parparita}, \bibinfo{person}{Alex Passos}, \bibinfo{person}{Mikhail Pavlov}, \bibinfo{person}{Andrew Peng}, \bibinfo{person}{Adam
  Perelman}, \bibinfo{person}{Filipe de Avila Belbute~Peres}, \bibinfo{person}{Michael Petrov}, \bibinfo{person}{Henrique~Ponde de Oliveira~Pinto}, \bibinfo{person}{Michael}, \bibinfo{person}{Pokorny}, \bibinfo{person}{Michelle Pokrass}, \bibinfo{person}{Vitchyr~H. Pong}, \bibinfo{person}{Tolly Powell}, \bibinfo{person}{Alethea Power}, \bibinfo{person}{Boris Power}, \bibinfo{person}{Elizabeth Proehl}, \bibinfo{person}{Raul Puri}, \bibinfo{person}{Alec Radford}, \bibinfo{person}{Jack Rae}, \bibinfo{person}{Aditya Ramesh}, \bibinfo{person}{Cameron Raymond}, \bibinfo{person}{Francis Real}, \bibinfo{person}{Kendra Rimbach}, \bibinfo{person}{Carl Ross}, \bibinfo{person}{Bob Rotsted}, \bibinfo{person}{Henri Roussez}, \bibinfo{person}{Nick Ryder}, \bibinfo{person}{Mario Saltarelli}, \bibinfo{person}{Ted Sanders}, \bibinfo{person}{Shibani Santurkar}, \bibinfo{person}{Girish Sastry}, \bibinfo{person}{Heather Schmidt}, \bibinfo{person}{David Schnurr}, \bibinfo{person}{John Schulman}, \bibinfo{person}{Daniel Selsam},
  \bibinfo{person}{Kyla Sheppard}, \bibinfo{person}{Toki Sherbakov}, \bibinfo{person}{Jessica Shieh}, \bibinfo{person}{Sarah Shoker}, \bibinfo{person}{Pranav Shyam}, \bibinfo{person}{Szymon Sidor}, \bibinfo{person}{Eric Sigler}, \bibinfo{person}{Maddie Simens}, \bibinfo{person}{Jordan Sitkin}, \bibinfo{person}{Katarina Slama}, \bibinfo{person}{Ian Sohl}, \bibinfo{person}{Benjamin Sokolowsky}, \bibinfo{person}{Yang Song}, \bibinfo{person}{Natalie Staudacher}, \bibinfo{person}{Felipe~Petroski Such}, \bibinfo{person}{Natalie Summers}, \bibinfo{person}{Ilya Sutskever}, \bibinfo{person}{Jie Tang}, \bibinfo{person}{Nikolas Tezak}, \bibinfo{person}{Madeleine~B. Thompson}, \bibinfo{person}{Phil Tillet}, \bibinfo{person}{Amin Tootoonchian}, \bibinfo{person}{Elizabeth Tseng}, \bibinfo{person}{Preston Tuggle}, \bibinfo{person}{Nick Turley}, \bibinfo{person}{Jerry Tworek}, \bibinfo{person}{Juan Felipe~Cerón Uribe}, \bibinfo{person}{Andrea Vallone}, \bibinfo{person}{Arun Vijayvergiya}, \bibinfo{person}{Chelsea Voss},
  \bibinfo{person}{Carroll Wainwright}, \bibinfo{person}{Justin~Jay Wang}, \bibinfo{person}{Alvin Wang}, \bibinfo{person}{Ben Wang}, \bibinfo{person}{Jonathan Ward}, \bibinfo{person}{Jason Wei}, \bibinfo{person}{CJ Weinmann}, \bibinfo{person}{Akila Welihinda}, \bibinfo{person}{Peter Welinder}, \bibinfo{person}{Jiayi Weng}, \bibinfo{person}{Lilian Weng}, \bibinfo{person}{Matt Wiethoff}, \bibinfo{person}{Dave Willner}, \bibinfo{person}{Clemens Winter}, \bibinfo{person}{Samuel Wolrich}, \bibinfo{person}{Hannah Wong}, \bibinfo{person}{Lauren Workman}, \bibinfo{person}{Sherwin Wu}, \bibinfo{person}{Jeff Wu}, \bibinfo{person}{Michael Wu}, \bibinfo{person}{Kai Xiao}, \bibinfo{person}{Tao Xu}, \bibinfo{person}{Sarah Yoo}, \bibinfo{person}{Kevin Yu}, \bibinfo{person}{Qiming Yuan}, \bibinfo{person}{Wojciech Zaremba}, \bibinfo{person}{Rowan Zellers}, \bibinfo{person}{Chong Zhang}, \bibinfo{person}{Marvin Zhang}, \bibinfo{person}{Shengjia Zhao}, \bibinfo{person}{Tianhao Zheng}, \bibinfo{person}{Juntang Zhuang},
  \bibinfo{person}{William Zhuk}, {and} \bibinfo{person}{Barret Zoph}.} \bibinfo{year}{2024}\natexlab{}.
\newblock \bibinfo{title}{GPT-4 Technical Report}.
\newblock
\newblock
\showeprint[arxiv]{2303.08774}~[cs.CL]


\bibitem[Palomba et~al\mbox{.}(2018)]%
        {10.1145/3180155.3182532}
\bibfield{author}{\bibinfo{person}{Fabio Palomba}, \bibinfo{person}{Gabriele Bavota}, \bibinfo{person}{Massimiliano Di~Penta}, \bibinfo{person}{Fausto Fasano}, \bibinfo{person}{Rocco Oliveto}, {and} \bibinfo{person}{Andrea De~Lucia}.} \bibinfo{year}{2018}\natexlab{}.
\newblock \showarticletitle{On the diffuseness and the impact on maintainability of code smells: a large scale empirical investigation}. In \bibinfo{booktitle}{\emph{Proceedings of the 40th International Conference on Software Engineering}} (Gothenburg, Sweden) \emph{(\bibinfo{series}{ICSE '18})}. \bibinfo{publisher}{Association for Computing Machinery}, \bibinfo{address}{New York, NY, USA}, \bibinfo{pages}{482}.
\newblock
\showISBNx{9781450356381}
\urldef\tempurl%
\url{https://doi.org/10.1145/3180155.3182532}
\showDOI{\tempurl}


\bibitem[Pantiuchina et~al\mbox{.}(2020)]%
        {10.1145/3408302}
\bibfield{author}{\bibinfo{person}{Jevgenija Pantiuchina}, \bibinfo{person}{Fiorella Zampetti}, \bibinfo{person}{Simone Scalabrino}, \bibinfo{person}{Valentina Piantadosi}, \bibinfo{person}{Rocco Oliveto}, \bibinfo{person}{Gabriele Bavota}, {and} \bibinfo{person}{Massimiliano~Di Penta}.} \bibinfo{year}{2020}\natexlab{}.
\newblock \showarticletitle{Why Developers Refactor Source Code: A Mining-based Study}.
\newblock \bibinfo{journal}{\emph{ACM Trans. Softw. Eng. Methodol.}} \bibinfo{volume}{29}, \bibinfo{number}{4}, Article \bibinfo{articleno}{29} (\bibinfo{date}{sep} \bibinfo{year}{2020}), \bibinfo{numpages}{30}~pages.
\newblock
\showISSN{1049-331X}
\urldef\tempurl%
\url{https://doi.org/10.1145/3408302}
\showDOI{\tempurl}


\bibitem[{Scientific Toolworks, Inc.}(2023)]%
        {understand_software}
\bibfield{author}{\bibinfo{person}{{Scientific Toolworks, Inc.}}} \bibinfo{year}{2023}\natexlab{}.
\newblock \bibinfo{title}{Understand}.
\newblock
\newblock
\urldef\tempurl%
\url{https://scitools.com/static-code-analysis}
\showURL{%
\tempurl}
\newblock
\shownote{Software for Static Code Analysis}.


\bibitem[scitools(2024)]%
        {understandmetrics}
\bibfield{author}{\bibinfo{person}{scitools}.} \bibinfo{year}{2024}\natexlab{}.
\newblock
\newblock
\urldef\tempurl%
\url{https://documentation.scitools.com/pdf/metricsdoc.pdf}
\showURL{%
\tempurl}


\bibitem[Sharma(2024a)]%
        {A_taxonomy_of_software_smells}
\bibfield{author}{\bibinfo{person}{Tushar Sharma}.} \bibinfo{year}{2024}\natexlab{a}.
\newblock
\newblock
\urldef\tempurl%
\url{https://tusharma.in/smells/CODE.html}
\showURL{%
\tempurl}


\bibitem[Sharma(2024b)]%
        {10.1145/3643991.3644881}
\bibfield{author}{\bibinfo{person}{Tushar Sharma}.} \bibinfo{year}{2024}\natexlab{b}.
\newblock \showarticletitle{Multi-faceted Code Smell Detection at Scale using DesigniteJava 2.0}. In \bibinfo{booktitle}{\emph{Proceedings of the 21st International Conference on Mining Software Repositories}} (Lisbon, Portugal) \emph{(\bibinfo{series}{MSR '24})}. \bibinfo{publisher}{Association for Computing Machinery}, \bibinfo{address}{New York, NY, USA}, \bibinfo{pages}{284–288}.
\newblock
\showISBNx{9798400705878}
\urldef\tempurl%
\url{https://doi.org/10.1145/3643991.3644881}
\showDOI{\tempurl}


\bibitem[Sharma et~al\mbox{.}(2016)]%
        {sharma2016designite}
\bibfield{author}{\bibinfo{person}{T. Sharma}, \bibinfo{person}{P. Mishra}, {and} \bibinfo{person}{R. Tiwari}.} \bibinfo{year}{2016}\natexlab{}.
\newblock \showarticletitle{Designite: A software design quality assessment tool}. In \bibinfo{booktitle}{\emph{Proceedings of the 1st International Workshop on Bringing Architectural Design Thinking into Developers’ Daily Activities}}. \bibinfo{pages}{1--4}.
\newblock


\bibitem[Shi et~al\mbox{.}(2023)]%
        {10.1145/3597926.3598036}
\bibfield{author}{\bibinfo{person}{Ensheng Shi}, \bibinfo{person}{Yanlin Wang}, \bibinfo{person}{Hongyu Zhang}, \bibinfo{person}{Lun Du}, \bibinfo{person}{Shi Han}, \bibinfo{person}{Dongmei Zhang}, {and} \bibinfo{person}{Hongbin Sun}.} \bibinfo{year}{2023}\natexlab{}.
\newblock \showarticletitle{Towards Efficient Fine-Tuning of Pre-trained Code Models: An Experimental Study and Beyond}. In \bibinfo{booktitle}{\emph{Proceedings of the 32nd ACM SIGSOFT International Symposium on Software Testing and Analysis}} (Seattle, WA, USA) \emph{(\bibinfo{series}{ISSTA 2023})}. \bibinfo{publisher}{Association for Computing Machinery}, \bibinfo{address}{New York, NY, USA}, \bibinfo{pages}{39–51}.
\newblock
\showISBNx{9798400702211}
\urldef\tempurl%
\url{https://doi.org/10.1145/3597926.3598036}
\showDOI{\tempurl}


\bibitem[Shirafuji et~al\mbox{.}(2023)]%
        {10479398}
\bibfield{author}{\bibinfo{person}{Atsushi Shirafuji}, \bibinfo{person}{Yusuke Oda}, \bibinfo{person}{Jun Suzuki}, \bibinfo{person}{Makoto Morishita}, {and} \bibinfo{person}{Yutaka Watanobe}.} \bibinfo{year}{2023}\natexlab{}.
\newblock \showarticletitle{Refactoring Programs Using Large Language Models with Few-Shot Examples}. In \bibinfo{booktitle}{\emph{2023 30th Asia-Pacific Software Engineering Conference (APSEC)}}. \bibinfo{pages}{151--160}.
\newblock
\urldef\tempurl%
\url{https://doi.org/10.1109/APSEC60848.2023.00025}
\showDOI{\tempurl}


\bibitem[Silva et~al\mbox{.}(2016)]%
        {10.1145/2950290.2950305}
\bibfield{author}{\bibinfo{person}{Danilo Silva}, \bibinfo{person}{Nikolaos Tsantalis}, {and} \bibinfo{person}{Marco~Tulio Valente}.} \bibinfo{year}{2016}\natexlab{}.
\newblock \showarticletitle{Why we refactor? confessions of GitHub contributors}. In \bibinfo{booktitle}{\emph{Proceedings of the 2016 24th ACM SIGSOFT International Symposium on Foundations of Software Engineering}} (Seattle, WA, USA) \emph{(\bibinfo{series}{FSE 2016})}. \bibinfo{publisher}{Association for Computing Machinery}, \bibinfo{address}{New York, NY, USA}, \bibinfo{pages}{858–870}.
\newblock
\showISBNx{9781450342186}
\urldef\tempurl%
\url{https://doi.org/10.1145/2950290.2950305}
\showDOI{\tempurl}


\bibitem[Svyatkovskiy et~al\mbox{.}(2020)]%
        {svyatkovskiy2020intellicode}
\bibfield{author}{\bibinfo{person}{Alexey Svyatkovskiy}, \bibinfo{person}{Shao~Kun Deng}, \bibinfo{person}{Shengyu Fu}, {and} \bibinfo{person}{Neel Sundaresan}.} \bibinfo{year}{2020}\natexlab{}.
\newblock \showarticletitle{Intellicode compose: Code generation using transformer}. In \bibinfo{booktitle}{\emph{Proceedings of the 28th ACM joint meeting on European software engineering conference and symposium on the foundations of software engineering}}. \bibinfo{pages}{1433--1443}.
\newblock


\bibitem[Touvron et~al\mbox{.}(2023)]%
        {touvron2023llamaopenefficientfoundation}
\bibfield{author}{\bibinfo{person}{Hugo Touvron}, \bibinfo{person}{Thibaut Lavril}, \bibinfo{person}{Gautier Izacard}, \bibinfo{person}{Xavier Martinet}, \bibinfo{person}{Marie-Anne Lachaux}, \bibinfo{person}{Timothée Lacroix}, \bibinfo{person}{Baptiste Rozière}, \bibinfo{person}{Naman Goyal}, \bibinfo{person}{Eric Hambro}, \bibinfo{person}{Faisal Azhar}, \bibinfo{person}{Aurelien Rodriguez}, \bibinfo{person}{Armand Joulin}, \bibinfo{person}{Edouard Grave}, {and} \bibinfo{person}{Guillaume Lample}.} \bibinfo{year}{2023}\natexlab{}.
\newblock \bibinfo{title}{LLaMA: Open and Efficient Foundation Language Models}.
\newblock
\newblock
\showeprint[arxiv]{2302.13971}~[cs.CL]
\urldef\tempurl%
\url{https://arxiv.org/abs/2302.13971}
\showURL{%
\tempurl}


\bibitem[Tsantalis et~al\mbox{.}(2020)]%
        {tsantalis2020refactoringminer}
\bibfield{author}{\bibinfo{person}{Nikolaos Tsantalis}, \bibinfo{person}{Ameya Ketkar}, {and} \bibinfo{person}{Danny Dig}.} \bibinfo{year}{2020}\natexlab{}.
\newblock \showarticletitle{RefactoringMiner 2.0}.
\newblock \bibinfo{journal}{\emph{IEEE Transactions on Software Engineering}} \bibinfo{volume}{48}, \bibinfo{number}{3} (\bibinfo{year}{2020}), \bibinfo{pages}{930--950}.
\newblock


\bibitem[Tsantalis et~al\mbox{.}(2018)]%
        {Tsantalis:ICSE:2018:RefactoringMiner}
\bibfield{author}{\bibinfo{person}{Nikolaos Tsantalis}, \bibinfo{person}{Matin Mansouri}, \bibinfo{person}{Laleh~M. Eshkevari}, \bibinfo{person}{Davood Mazinanian}, {and} \bibinfo{person}{Danny Dig}.} \bibinfo{year}{2018}\natexlab{}.
\newblock \showarticletitle{Accurate and Efficient Refactoring Detection in Commit History}. In \bibinfo{booktitle}{\emph{Proceedings of the 40th International Conference on Software Engineering}} (Gothenburg, Sweden) \emph{(\bibinfo{series}{ICSE '18})}. \bibinfo{publisher}{ACM}, \bibinfo{address}{New York, NY, USA}, \bibinfo{pages}{483--494}.
\newblock
\showISBNx{978-1-4503-5638-1}
\urldef\tempurl%
\url{https://doi.org/10.1145/3180155.3180206}
\showDOI{\tempurl}


\bibitem[Van~Emden and Moonen(2002)]%
        {van2002java}
\bibfield{author}{\bibinfo{person}{Eva Van~Emden} {and} \bibinfo{person}{Leon Moonen}.} \bibinfo{year}{2002}\natexlab{}.
\newblock \showarticletitle{Java quality assurance by detecting code smells}. In \bibinfo{booktitle}{\emph{Ninth Working Conference on Reverse Engineering, 2002. Proceedings.}} IEEE, \bibinfo{pages}{97--106}.
\newblock


\bibitem[Wei et~al\mbox{.}(2023)]%
        {wei2023copiloting}
\bibfield{author}{\bibinfo{person}{Yuxiang Wei}, \bibinfo{person}{Chunqiu~Steven Xia}, {and} \bibinfo{person}{Lingming Zhang}.} \bibinfo{year}{2023}\natexlab{}.
\newblock \showarticletitle{Copiloting the copilots: Fusing large language models with completion engines for automated program repair}. In \bibinfo{booktitle}{\emph{Proceedings of the 31st ACM Joint European Software Engineering Conference and Symposium on the Foundations of Software Engineering}}. \bibinfo{pages}{172--184}.
\newblock


\bibitem[Xu et~al\mbox{.}(2022a)]%
        {xu2022systematic}
\bibfield{author}{\bibinfo{person}{Frank~F Xu}, \bibinfo{person}{Uri Alon}, \bibinfo{person}{Graham Neubig}, {and} \bibinfo{person}{Vincent~Josua Hellendoorn}.} \bibinfo{year}{2022}\natexlab{a}.
\newblock \showarticletitle{A systematic evaluation of large language models of code}. In \bibinfo{booktitle}{\emph{Proceedings of the 6th ACM SIGPLAN International Symposium on Machine Programming}}. \bibinfo{pages}{1--10}.
\newblock


\bibitem[Xu et~al\mbox{.}(2022b)]%
        {10.1145/3520312.3534862}
\bibfield{author}{\bibinfo{person}{Frank~F. Xu}, \bibinfo{person}{Uri Alon}, \bibinfo{person}{Graham Neubig}, {and} \bibinfo{person}{Vincent~Josua Hellendoorn}.} \bibinfo{year}{2022}\natexlab{b}.
\newblock \showarticletitle{A systematic evaluation of large language models of code}. In \bibinfo{booktitle}{\emph{Proceedings of the 6th ACM SIGPLAN International Symposium on Machine Programming}} (San Diego, CA, USA) \emph{(\bibinfo{series}{MAPS 2022})}. \bibinfo{publisher}{Association for Computing Machinery}, \bibinfo{address}{New York, NY, USA}, \bibinfo{pages}{1–10}.
\newblock
\showISBNx{9781450392730}
\urldef\tempurl%
\url{https://doi.org/10.1145/3520312.3534862}
\showDOI{\tempurl}


\bibitem[Xu et~al\mbox{.}(2022c)]%
        {10.1145/3487569}
\bibfield{author}{\bibinfo{person}{Frank~F. Xu}, \bibinfo{person}{Bogdan Vasilescu}, {and} \bibinfo{person}{Graham Neubig}.} \bibinfo{year}{2022}\natexlab{c}.
\newblock \showarticletitle{In-IDE Code Generation from Natural Language: Promise and Challenges}.
\newblock \bibinfo{journal}{\emph{ACM Trans. Softw. Eng. Methodol.}} \bibinfo{volume}{31}, \bibinfo{number}{2}, Article \bibinfo{articleno}{29} (\bibinfo{date}{mar} \bibinfo{year}{2022}), \bibinfo{numpages}{47}~pages.
\newblock
\showISSN{1049-331X}
\urldef\tempurl%
\url{https://doi.org/10.1145/3487569}
\showDOI{\tempurl}


\bibitem[Zamfirescu-Pereira et~al\mbox{.}(2023)]%
        {10.1145/3544548.3581388}
\bibfield{author}{\bibinfo{person}{J.D. Zamfirescu-Pereira}, \bibinfo{person}{Richmond~Y. Wong}, \bibinfo{person}{Bjoern Hartmann}, {and} \bibinfo{person}{Qian Yang}.} \bibinfo{year}{2023}\natexlab{}.
\newblock \showarticletitle{Why Johnny Can’t Prompt: How Non-AI Experts Try (and Fail) to Design LLM Prompts}. In \bibinfo{booktitle}{\emph{Proceedings of the 2023 CHI Conference on Human Factors in Computing Systems}} (, Hamburg, Germany,) \emph{(\bibinfo{series}{CHI '23})}. \bibinfo{publisher}{Association for Computing Machinery}, \bibinfo{address}{New York, NY, USA}, Article \bibinfo{articleno}{437}, \bibinfo{numpages}{21}~pages.
\newblock
\showISBNx{9781450394215}
\urldef\tempurl%
\url{https://doi.org/10.1145/3544548.3581388}
\showDOI{\tempurl}


\bibitem[Zheng et~al\mbox{.}(2023)]%
        {10.1145/3580305.3599790}
\bibfield{author}{\bibinfo{person}{Qinkai Zheng}, \bibinfo{person}{Xiao Xia}, \bibinfo{person}{Xu Zou}, \bibinfo{person}{Yuxiao Dong}, \bibinfo{person}{Shan Wang}, \bibinfo{person}{Yufei Xue}, \bibinfo{person}{Lei Shen}, \bibinfo{person}{Zihan Wang}, \bibinfo{person}{Andi Wang}, \bibinfo{person}{Yang Li}, \bibinfo{person}{Teng Su}, \bibinfo{person}{Zhilin Yang}, {and} \bibinfo{person}{Jie Tang}.} \bibinfo{year}{2023}\natexlab{}.
\newblock \showarticletitle{CodeGeeX: A Pre-Trained Model for Code Generation with Multilingual Benchmarking on HumanEval-X}. In \bibinfo{booktitle}{\emph{Proceedings of the 29th ACM SIGKDD Conference on Knowledge Discovery and Data Mining}} (, Long Beach, CA, USA,) \emph{(\bibinfo{series}{KDD '23})}. \bibinfo{publisher}{Association for Computing Machinery}, \bibinfo{address}{New York, NY, USA}, \bibinfo{pages}{5673–5684}.
\newblock
\showISBNx{9798400701030}
\urldef\tempurl%
\url{https://doi.org/10.1145/3580305.3599790}
\showDOI{\tempurl}


\end{thebibliography}

\appendix

\end{document}